\documentclass[
 prx,
 twocolumn,
 10pt,
 superscriptaddress,
 amsmath,
 amssymb,
 aps,
 xcolor,
 tabularx,
 longbibliography
]{revtex4-2}

\usepackage{bm}
\usepackage{xcolor}          %
\usepackage{graphicx}
\usepackage[letterpaper,margin=1in]{geometry}
\usepackage{url}
\usepackage{makecell}
\usepackage{multirow}
\usepackage{tabularray}
\usepackage{physics}
\usepackage[normalem]{ulem}  %
\usepackage{xparse}
\usepackage{soul} %
\usepackage{cancel}

\newcommand{\mc}{\mathcal}

\newcommand{\mr}{\mathrm}
\newcommand{\mf}{\mathfrak}

\newcommand{\mbb}{\mathbb}
\newcommand{\pt}{\mathrm{pt}}
\newcommand{\Pn}{{\mathcal P}^n}
\newcommand{\E}{{\mathop{\mbb{E}}}}
\newcommand{\lket}[1]{\vert #1 \rangle\!\rangle}
\newcommand{\lbra}[1]{\langle\!\langle #1 \vert}

\newcommand{\lketbra}[2]{\vert #1 \rangle\!\rangle\langle\!\langle #2 \vert}

\setstcolor{blue}

\DeclareRobustCommand{\remove}[2][]{%
  \begingroup
    \def\ULthickness{0.5pt}%
    \if\relax\detokenize{#1}\relax
      \textcolor{red}{\sout{#2}}%
    \else
      \textcolor{#1}{\sout{#2}}%
    \fi
  \endgroup
}

\usepackage{hyperref}
\hypersetup{
    urlcolor=blue,
    citecolor=blue,
    colorlinks=true,
    linkcolor=blue
}

\newtheorem{theorem}{Theorem}
\newtheorem{proposition}[theorem]{Proposition}%

\begin{document} 
\title{Disambiguating Pauli noise in quantum computers}
\author{Edward~H.~Chen}
\thanks{Co-leading authors listed  in  alphabetical order.}
\affiliation{IBM Quantum, Research Triangle Park, North Carolina. 27709, USA}

\author{Senrui~Chen}
\thanks{Co-leading authors listed  in  alphabetical order.}
\affiliation{Pritzker School of Molecular Engineering, University of Chicago, Chicago  60637, USA}

\author{Laurin~E.~Fischer}
\thanks{Co-leading authors listed  in  alphabetical order.}
\affiliation{IBM Quantum, IBM Research Europe – Z{\"u}rich, 8803 R{\"u}schlikon, Switzerland}
\affiliation{Theory and Simulation of Materials, {\'E}cole Polytechnique F{\'e}d{\'e}rale de Lausanne, 1015 Lausanne, Switzerland}
\author{Andrew~Eddins}
\affiliation{IBM Quantum, Almaden Research Center, San Jose, 95120, USA}
\author{Luke~C.~G.~Govia}
\affiliation{IBM Quantum, Almaden Research Center, San Jose, 95120, USA}
\author{Brad~Mitchell}
\affiliation{IBM Quantum, Almaden Research Center, San Jose, 95120, USA}
\author{Andre~He}
\affiliation{IBM Quantum, T. J. Watson Research Center, Yorktown Heights, 10598, USA}
\author{Youngseok~Kim}
\affiliation{IBM Quantum, T. J. Watson Research Center, Yorktown Heights, 10598, USA}
\author{Liang~Jiang}
\affiliation{Pritzker School of Molecular Engineering, University of Chicago, Chicago  60637, USA}
\author{Alireza~Seif}
\affiliation{IBM Quantum, T. J. Watson Research Center, Yorktown Heights, 10598, USA}

\begin{abstract} 
To successfully perform quantum computations, it is often necessary to first accurately characterize the noise in the underlying hardware. However, it is well known that fundamental limitations prevent the unique identification of the noise model. This raises the question of whether these limitations impact the ability to predict noisy dynamics and mitigate errors. Here, we show, both theoretically and experimentally, that when learnable parameters are self-consistently characterized, the \textit{unlearnable} (gauge) degrees of freedom do not impact predictions of noisy dynamics or error mitigation. We use the recently introduced framework of gate set Pauli noise learning to efficiently and self-consistently characterize and mitigate noise of a complete gate set, including state preparation, measurements, single-qubit gates and multi-qubit entangling Clifford gates. We validate our approach through experiments with up to 92 qubits and show that, while the gauge choice does not affect error-mitigated observables, optimizing it reduces sampling overhead
\end{abstract}

\maketitle
\section{INTRODUCTION}
Quantum computers are believed to be exponentially faster than classical computers for many important problems~\cite{QuantumAlgorithmZoo}.
However, noise limits the performance of the quantum hardware, motivating the widespread efforts to characterize the noise in order to address it. Important areas where noise learning protocols are expected to have on-going impact include: quantifying improvements to hardware architectures~\cite{mckay2023benchmarkingquantumprocessorperformance}, mitigating the impact of noise on observables with additional quantum and classical processing~\cite{kim2023evidence}, or improving algorithms needed to actively correct noise soon after it occurs~\cite{PhysRevLett.128.110504}. As progress is made along all directions, it is increasingly accepted that quantum computations will also \textit{continuously}, as opposed to abruptly, improve in accuracy~\cite{cai2023quantum, 10.1063/5.0082975}.

\begin{figure*}[t]
    \centering
	\includegraphics[width=1.0\textwidth]{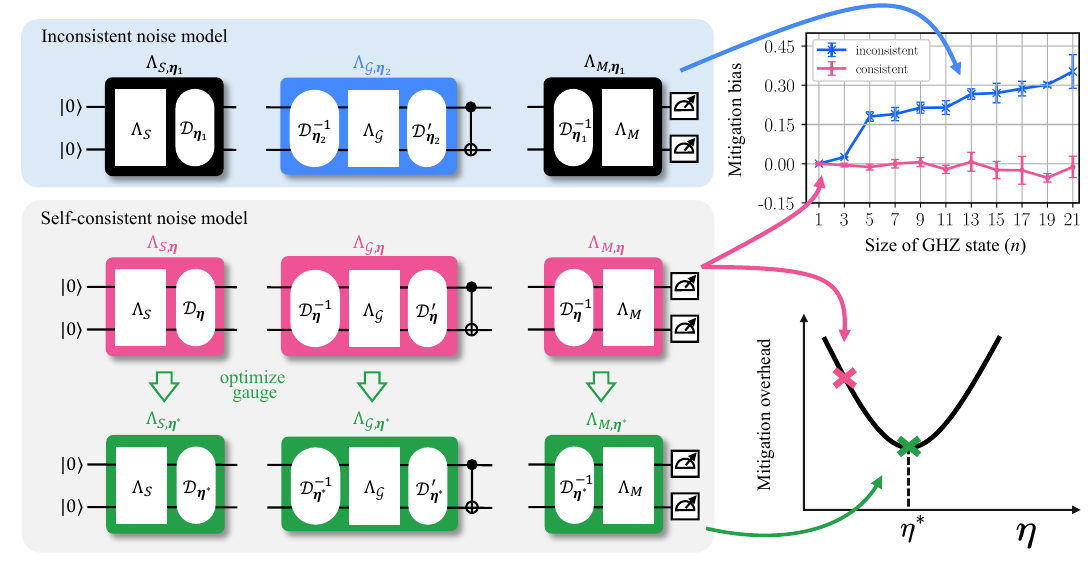}
	\caption[Overview of results.]{
		\textbf{Overview of results.} 
		Leading error mitigation methods based on Pauli noise models presuppose accurate knowledge of all the hardware error rates. 
        This foundational assumption, however, is false, as it has been proven that such a noise model cannot be uniquely determined by experiments, even in principle. 
        Without accounting for this indeterminacy, previous experiments implicitly used an \textit{inconsistent} set of the gauge parameters $\{\mathcal{D}_{\boldsymbol{\eta}_1}, \mathcal{D}_{\boldsymbol{\eta}_2} \}$ across the quantum gate set, e.g., different gauge choices for state preparation and measurement and the two-qubit gate (top, blue).
        We show that a self-consistent set of gauge parameters (middle, pink) is necessary for unbiased quantum error mitigation, as exemplified here in the mitigation bias of state preparation experiment of an $n=21$ entangled state known as the Greenberger–Horne–Zeilinger state (GHZ) state (upper right; details in Fig.~\ref{fig:6}).
        Furthermore, the choice of a consistent gauge can be optimized (bottom, green) to reduce the sampling overhead of error mitigation (bottom, right). 
	}
	\label{fig:1}
\end{figure*}

Recent progress towards building larger quantum computers has highlighted a need for scalable methods to fully characterize all possible types of quantum noise, which can be as intractable to classically model as the quantum algorithm being executed~\cite{lidar2013introduction}. 
To reduce the complexity of the learning task, the predominant sources of noise are assumed to only impact the qubit subspace, and are also physically localized to neighboring qubits on the device.
Upon transforming the underlying noise using randomized compiling or Pauli twirling~\cite{wallman2016noise}, a Pauli noise model becomes a practical choice because it can be made as complex as necessary while remaining classically tractable~\cite{cai2023quantum}. 
In fact, it was recently shown that a learned noise model could be used to effectively mitigate noise in applications which require accurate estimates of expectation values~\cite{van2023probabilistic, kim2023evidence, fischer2024dynamical}.

Such error mitigation strategies can in principle yield unbiased estimators at the cost of additional quantum circuit executions (shots), with the assumption that the device noise is faithfully captured by the learned noise model~\cite{temme2017error}. The original proposal of Ref.~\cite{endo2018practical} addressed this requirement by building on gate set tomography (GST)~\cite{nielsen2021gate}, which learns gate, state-preparation, and measurement noise self-consistently and naturally accounts for gauge degrees of freedom in the noise description. However, error mitigation with GST has been limited in practice to one- and two-qubit systems due to its unfavorable scaling~\cite{song2019quantum,zhang2020error}, with even three-qubit GST
requiring considerable experimental
effort~\cite{madzik2022precision}. Therefore, subsequent work turned to more scalable noise-learning approaches such as cycle benchmarking~\cite{erhard2019characterizing, calzona2024multi} and sparse Pauli-Lindblad models~\cite{van2023probabilistic, van2022model, kim2023evidence} to enable error mitigation on processors with tens to over a hundred qubits. These scalable methods achieve their efficiency by characterizing gate noise using SPAM-robust techniques and treating SPAM errors separately. While practically convenient, this separation comes at a cost. It has recently been shown that certain combinations of gate and SPAM noise parameters cannot be uniquely identified when learned independently, giving rise to gauge degrees of freedom in the Pauli noise model~\cite{chen2023learnability, chen2026efficient}. When SPAM and gate noise are characterized separately, these gauge-ambiguous combinations cannot be correctly resolved, which leads to inconsistencies in the predictions of the error model (see Fig.~\ref{fig:1}).

Although the existence of such gauge ambiguities has been established theoretically~\cite{chen2026efficient}, their practical consequences for error mitigation have not been investigated. It has remained an open question how large these inconsistencies are on real hardware and whether they meaningfully degrade mitigation performance. Here, we address both questions. We prove theoretically that by self-consistently inferring all the \textit{learnable} parameters in a Pauli noise model, including SPAM and gate noise together, it is possible to predict the outcomes of any noisy experiment and successfully perform error mitigation, even without resolving the fundamentally unlearnable gauge parameters. We further provide extensive experimental evidence, quantifying the magnitude of gauge-induced inconsistencies in practical error mitigation and demonstrating that self-consistent learning eliminates them. The key enabler is the Pauli gate set learning method of Ref.~\cite{chen2026efficient}, which treats SPAM and gate noise within a unified framework similar in spirit to GST~\cite{nielsen2021gate} but specialized to Pauli noise for scalability and efficiency. Our experiments use this framework to identify the learnable parameters of Pauli noise channels and to \textit{unambiguously} and efficiently characterize them. We review this framework in Sec.~\ref{sec:model_learning} and discuss how gauge degrees of freedom emerge as a result of SPAM errors. We then discuss the application of this framework to a quasi-local noise model in Sec.~\ref{sec:local}.

In Sec.~\ref{sec:theory_validation}, we show that such noise models naturally enable a self-consistent and unbiased error mitigation strategy. Specifically, we show that when probabilistic error cancellation (PEC)~\cite{van2023probabilistic} is implemented with self-consistently learned noise models and applied to SPAM and gate errors, it produces unbiased estimates of observables that do not depend on the gauge degrees of freedom. 
We further show, surprisingly, that despite not impacting observables, changing the gauge parameters does impact the overhead of required shots for error mitigation.
Building on this insight, we propose and demonstrate a scalable method for identifying the gauge parameters needed to minimize this sampling overhead.

We demonstrate the learning framework and our theoretical results by performing several error mitigation experiments with increasing complexity, and show that it reduces the bias in mitigated expectation values compared to previous approaches. Specifically, we start from a simple two-qubit example in Sec.~\ref{sec:2qubit} and show that inconsistencies in handling gauge degrees of freedom in previous error mitigation techniques lead to errors in mitigated expectation values, whereas our method provides consistent and accurate estimates. Building on these results, we next consider mitigating expectation values of high-weight stabilizers of Greenberger-Horn-Zeilinger (GHZ)~\cite{greenberger1989going} states on up to 21 qubits in Sec.~\ref{sec:GHZ}. 
In these experiments, we do not impose locality on the error model, but instead rely on the stabilizer nature of the target state to simplify the mitigation by only learning a subset of error parameters. 
We again observe that while inconsistencies limit the accuracy of previously used methods, our method succeeds in producing correct error-mitigated estimates.
To showcase the scalability of the learning protocol, we consider brickwork circuits on a ring of 92 qubits, and learn the full quasi-local noise model in Sec.~\ref{sec:ring}. We consider 92 single-qubit observables and again observe that our method generally reduces bias compared to previous techniques. 
Finally, in Sec.~\ref{sec:gauge-optimized_PEC} we perform PEC experiments on non-Clifford circuits of 20 qubits which confirm that optimizing the gauge of a self-consistently learned model can significantly reduce the shot overhead of error mitigation. See overview of all the sections in Fig.~\ref{fig:1}.

\section{MATHEMATICAL PRELIMINARIES}

\subsection{\label{sec:model_learning}Modeling and learning a gate set}
A Pauli channel on $n$ qubits is a stochastic mixture of $n$-qubit Pauli operators $P_a \in \mathcal{P}^n = \{I,X,Y,Z\}^{\otimes n}$ described by a $4^n$-dimensional probability distribution $\{p_a\}$, known as the Pauli error rates. One property of Pauli channels is that they transform any Pauli operator $P_a$ to itself up to a prefactor $\lambda_a\in[-1,1]$, known as the Pauli eigenvalues. Mathematically, a Pauli channel can be represented in the following two ways,
\begin{equation}
    \Lambda(\rho)=\sum_{a\in\Pn}p_a P_a\rho P_a = \frac1{2^n}\sum_{b\in\Pn}\lambda_bP_b\mr{tr}(P_b\rho).
\end{equation}
Both representations have $4^n-1$ degrees of freedom, as the trace-preserving condition of quantum channels requires $\sum_a p_a=1$, or equivalently $\lambda_{I^{\otimes n}}=1$. 
For now, we consider Pauli channels that are completely general. We will discuss Pauli channels with efficient parameterization (e.g., quasi-local Pauli channels) in the next subsection.

Our work considers a ``gate set''~\cite{nielsen2021gate}
comprised of state preparation, measurement, single-qubit gates, and entangling gates (See Fig.~\ref{fig:1}).
Concretely, let the ideal initial state be $\rho_0=\ketbra{0}{0}^{\otimes n}$, the measurement be the projection onto the computational basis $\mc M_Z$, the entangling gates be a finite collection of Clifford gates $\{\mc G\}$, and the single-qubit gates be arbitrary $\{\mc U=\bigotimes_{i=1}^n\mc U_i\}$.
In practice, the gate set is noisy. We use a Pauli noise model to describe the noisy gate set, where state preparation, measurement, and entangling gates are subject to Pauli noise channels,
\begin{equation}
    \tilde{\rho}_0=\Lambda^S(\rho_0),\quad \tilde{\mc M}_Z=\mc M_Z\circ\Lambda^M,\quad \tilde{\mc G} = \mc G\circ\Lambda^{\mc G}.
\end{equation}
We further assume that the single-qubit gates have negligible noise (which can be relaxed to gate-independent noise~\cite{wallman2016noise}),
and that the SPAM noise channels $\Lambda^S$ and $\Lambda^M$ are generalized depolarizing channels, which are Pauli channels whose Pauli error rates only depend on the support of the corresponding Pauli operators, and thus contain only $2^n-1$ degrees of freedom.
These assumptions about the noise channels can be physically enforced using randomized compiling or Pauli twirling given reasonably good single-qubit control, as have been widely adopted and verified in the literature~\cite{wallman2016noise, hashim2020randomized,van2023probabilistic, ferracin2024efficiently}.

Before we discuss how to learn the noisy gate set, it is crucial to note that not every noise parameter is identifiable~\cite{huang2022foundations,chen2023learnability,chen2026efficient}. To see this, we highlight the fact that for any quantum circuit and any observable, the noisy expectation value takes the following form,
\begin{equation}\label{eq:expression_expectation}
\begin{split}
     \langle\tilde o\rangle &= \sum_{a_0,\cdots,a_{T+1}\in\mc P^n}c_{\bm a} \lambda_{a_0}^S\lambda_{a_1}^{\mc G_1}\cdots\lambda_{a_T}^{\mc G_T}\lambda_{a_{T+1}}^M\\ &= \sum_{a_0,\cdots,a_{T+1}\in\mc P^n} c_{\bm a}\Gamma_{\bm a}.
\end{split}
\end{equation}
Here, there are $T$ layers of entangling Clifford gates $\mc G_1,\cdots,\mc G_T$ in the circuits, possibly interleaved by single-qubit gates. $\{c_{\bm a}\}$ are real numbers depending only on the ideal circuits and the observables, but not on the noise parameters. $\Gamma_{\bm a}=\lambda_{a_0}^S\lambda_{a_1}^{\mc G_1}\cdots\lambda_{a_T}^{\mc G_T}\lambda_{a_{T+1}}^M$ is a product of Pauli eigenvalues known as a Pauli path~\cite{aharonov2023polynomial}.
\begin{figure}[b]
    \centering
	\includegraphics[width=0.35\textwidth]{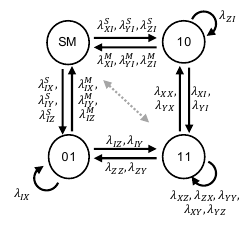}
	\caption{
		\textbf{Pattern transfer graph}. Example of a $2$-qubit gate set that contains CNOT as the only entangling gate, where any observable can be mapped to a cycle starting from the State-preparation and Measurement (SM) node and back. The bi-directional gray dashed line has 18 SPAM eigenvalues associated with it that have been omitted for clarity. The numbers in the nodes denote \emph{patterns} of Pauli operators, i.e., a bit string indicating which qubits have a nontrivial single-qubit Pauli operator supported on them.}
	\label{fig:2} 
\end{figure}

Importantly, $\Gamma_a$ cannot be arbitrary monomials of Pauli eigenvalues. The allowed set of $\Gamma_a$ can be described using a directed graph called the pattern transfer graph~\cite{chen2023learnability,chen2026efficient}, which describes how the gate set transform Pauli operators. 
The \emph{pattern} of an $n$-qubit Pauli operator $a$ is an $n$-bit string, denoted by $\pt(a)$, whose $i$th bit is $1$ if $a_i$ is a nontrivial Pauli and $0$ otherwise. For example, $\pt(IXIYZ)=01011$. 
See Fig.~\ref{fig:2} for an example of a pattern transfer graph for a 2-qubit system with Controlled-Not (CNOT) being the only entangling Clifford gate between a control qubit (left index) and a target qubit (right index), e.g. CNOT(IZ) $\rightarrow$ CNOT(ZZ).
Each edge on the pattern transfer graph corresponds to a unique Pauli eigenvalue from one of the Pauli noise channels.
Any path on the graph corresponds to a product of Pauli eigenvalues along the path.
It is known that the set of valid $\{\Gamma_{\bm a}\}$ has a one-to-one correspondence with the set of paths starting from and ending at the root node denoted as ``SM''~\cite{chen2023learnability,chen2026efficient}.
Consequently, the products of Pauli eigenvalues on cycles completely determine the outcomes of all possible experiments within the noisy gate set. 
If we transform the Pauli eigenvalues in a way that preserves the value of all cycles, then no experiments can witness such a transformation, which means there are gauge (i.e., non-identifiable) degrees of freedom. It was shown in Ref.~\cite{chen2026efficient} that all gauge transformations in the Pauli noise model can be expressed as
\begin{align}\label{eq:gauge_trans}
    &\Lambda^S\mapsto\Lambda_{\bm\eta}^S=\mc D_{\bm\eta}\circ\Lambda^S,\\ &\Lambda^M\mapsto\Lambda^M_{\bm\eta}=\Lambda^M\circ\mc D_{\bm\eta}^{-1},\\ & \Lambda^{\mc G}\mapsto\Lambda^{\mc G}_{\bm\eta}= \mc D_{\bm\eta}'\circ\Lambda^{\mc G}\circ\mc D_{\bm\eta}^{-1}.
\end{align}
Here, $\mc D_{\bm\eta}$ is any generalized depolarizing map, written as
\begin{equation}\label{eq:generalized_dep}
    \mc D_{\bm\eta}(\rho)=\sum_{a\in\Pn}e^{-\eta_{\mr{pt}(a)}} P_a\tr(P_a\rho)/2^n.
\end{equation}
with a real vector $\bm\eta$ we refer to as the gauge parameters where $\eta_{0_n}=0$ by the trace-preserving condition. 

Recall that $\pt(a)$ is the pattern of the the Pauli operator $a$.

$\mc D_{\bm\eta}'$ is defined by $\mc D_{\bm\eta}' = \mc G^{-1}\circ\mc D_{\bm\eta}\circ\mc G$. Note that $\mc D_{\bm\eta}$ commutes with any single-qubit gates. Thus, the transformations in Eq.~\eqref{eq:gauge_trans} preserve any experimental outcome, and also all noise assumptions of the Pauli noise model. This is illustrated in Fig.~\ref{fig:1}.
The remaining consideration is the positivity of the transformed channels -- excluding Pauli channels on the boundary of the set of positive maps, any sufficiently small $\bm\eta$ yield physical channels~\cite{chen2023learnability}.
Furthermore, in applications like error mitigation to be discussed later, it is acceptable to work with $\Lambda_{\bm\eta}$ that are not positive. 
Thus, a noisy gate set can be learned up to the $2^n-1$ gauge parameters parameterized by $\bm\eta$~\cite{chen2026efficient}.

To learn the gate set self-consistently, we first define the logarithm of the Pauli eigenvalues $x_a = -\log\lambda_a$ for all the Pauli channels $\{\Lambda^S,\Lambda^M,\{\Lambda^{\mc G}\}\}$. 
Let $\bm x$ be a vector comprised of all $\{x_a\}$, the length of which depends on the size of the gate set and the number of qubits.
We will design a set of experiments to learn $\bm x$, where each experiment consists of a sequence of Clifford gates and a Pauli observable measured at the end, the expectation value of which satisfies $\langle\tilde o\rangle = \Gamma_{\bm a}$. Taking the negative logarithm on both sides yields,
\begin{equation}
    -\log\langle\tilde o\rangle = x_{a_0}^S+x_{a_1}^{\mc G_1}+\cdots+x_{a_T}^{\mc G_T}+x_{a_{T+1}}^M, 
\end{equation}
which is a linear equation of $\bm x$. We emphasize that unlike Eqn.~\eqref{eq:expression_expectation} which holds for \textit{any} observable for any general circuit, this expression requires only a monomial expression because it refers to a Pauli observable for a Clifford circuit. Combining the linear equations from all experiments, we arrive at
\begin{equation}\label{eq:design_matrix}
    \bm b = F\bm x,
\end{equation}
where $b_j=-\log{\expval{\tilde o_j}}$ is the (log) expectation value for the $j$-th measured Pauli observable on the $j$-th circuit, and $F$ is called the design matrix. Our first goal is to collect enough experiments such that $F$ has the maximal possible rank. That is, the dimension of the null space of $F$ equals the number of gauge parameters, $2^n-1$. 
This can be achieved by including some experiments that contain no entangling gates (called depth-$0$ experiments, $b=x_{a_0}^S+x_{a_1}^M$) or contain one layer of entangling gates (called depth-$1$ experiments, $b=x_{a_0}^S+x^{\mc G}_{a_1}+x_{a_2}^M$);
to improve the estimated precision of model parameters, we can also include experiments that concatenate multiple layers of entangling gates (e.g., depth-$k$ experiments, $b=x_{10}^S+k\,x^{\mr{CNOT}}_{ZI}+x_{10}^M$).
similar to cycle error reconstruction~\cite{erhard2019characterizing,carignan2023error,flammia2020efficient}. More details about the experimental construction are presented in Sec.~\ref{supp:constructing_design_matrix}.

In practice, the experiments specified by $F$ are each run many times to obtain an estimate for the vector ${\bm b}$. 
Multiple rows $\{ j \}$ of $F$ may be estimated from one experimental setting, provided the Paulis $\{ o_j \}$ are site-wise commuting and the circuits are the same.
Minimizing the residual error in the least-squares problem $\|F\bm{x}_0-\bm b\|\le\epsilon$, where $\epsilon$ is chosen based on the tolerable amount of residual errors, produces a solution $\bm x_0$.
The final estimate yields $\bm x_{\bm\eta}=\bm x_0 + \bm y_{\bm\eta}$, where $\bm y_{\bm\eta}$ is a gauge vector depending on the gauge parameters $\bm\eta$, associated with the kernel of $F$.

The key difference between this approach and previous attempts at Pauli noise learning~\cite{flammia2020efficient,erhard2019characterizing,van2023probabilistic,carignan2023error,van2024techniques} is that we consider the full gate set, as opposed to only subsets of it. Though each Pauli noise channel can only be determined up to a gauge transformation, they are related by the same gauge parameters $\bm\eta$.
Our approach resembles a technique known as averaged circuit eigenvalue sampling (or ACES)~\cite{flammia:LIPIcs.TQC.2022.4,hockings2025scalable,hockings2025improving,pelaez2024average} which solves a system of linear equations similar to Eq.~\eqref{eq:design_matrix}. A key difference is, whereas ACES constructs a full-rank design matrix by introducing additional assumptions and may not learn every learnable parameter, our design matrix fully characterizes all learnable parameters, leaving only the gauge undetermined.

\subsection{Quasi-local noise models \label{sec:local}}
\begin{figure}[b]
    \centering
	\includegraphics[width=0.5\textwidth]{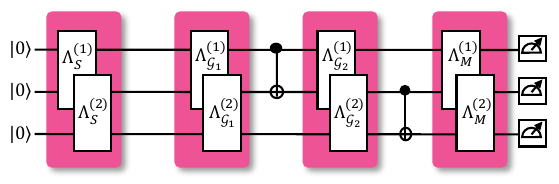}
	\caption{
		\textbf{Quasi-local noise model.} An example of a $3$-qubit system with $2$-local noise generators.
	}
	\label{fig:3}
\end{figure}

The above framework can apply to larger systems by imposing a quasi-local noise model, such that the Pauli noise channels are determined by a number of parameters linear in the number of qubits. Such underlying assumptions about the locality of the noise are supported by experimental successes to date~\cite{flammia2020efficient,harper2020efficient,harper2023learning,van2023probabilistic,wagner2023learning,rouze2023efficient,chen2026efficient, fischer2024dynamical}
A quasi-local Pauli noise channel is defined as 
\begin{equation}\label{eq:quasi_local}
    \Lambda(\cdot)=\mathop{\bigcirc}_{a\in\mathcal{K}}(\omega_a P_a(\cdot) P_a+(1-\omega_a)(\cdot)),
\end{equation}
where $\bigcirc$ denotes composition of maps, and $\mathcal{K}$ is a set of local Pauli operators; that is, they are supported on a local subset of qubits. The ordering in this decomposition does not matter because the Pauli channels commute. Equivalently, we can define the generator of the channel $\mathcal{L}$ such that $\Lambda=e^\mathcal{L}$, where 
\begin{equation}\label{eq:paulilind}
    \mathcal{L}(\rho) = \sum_{a\in\mathcal{K}} \frac{\tau_a}{2} (P_a\rho P_a-\rho).
\end{equation}
We require $\omega_a<1/2$ and define $\tau_a = -\log(1-2\omega_a)$ as the generator rate of the channel. 
Note that we allow $\tau_a$ (and thus $\omega_a$) to be negative.
We can then map the generator rates to log-fidelities through
\begin{equation}\label{eq:model_to_lind}
\begin{aligned}
        x_a &= \sum_{b\in\mathcal{K}}\langle{a,b}\rangle\tau_b,&&\forall a\in\Pn,\,a\neq0,
\end{aligned}
\end{equation}

where $\langle a,b\rangle=0$ if $P_a$ commutes with $P_b$ and $1$ otherwise.  For this relationship to hold, the set of local operators $\mathcal{K}$ should satisfy certain mathematical properties as explained in Sec.~\ref{supp:quasi-model}. 
Another equivalent understanding is that a quasi-local Pauli channel can be expressed as a composition of (possibly non-positive) Pauli channels supported on local subsystems, as shown in Fig.~\ref{fig:3}.
The noise parameters are given by $\bm\tau$, a vector comprised of all generators $\{\tau_a\}$ from each noise channel. 
Equivalently, the noise parameters can be chosen as $\bm x$ containing only $x_a$ such that $a\in\mc K$, which is related to $\bm\tau$ by an invertible matrix as in Eq.~\eqref{eq:model_to_lind}.
Another useful way to parameterize a quasi-local Pauli channel is through the M\"obius inversion of $\bm x$~\cite{chen2026efficient,wagner2023learning}, denoted by $\bm r$, which is a vector of length $|\mc K|$ interchangeable with $\bm\tau$ by an invertible matrix. We define $\bm r$ in Sec.~\ref{supp:quasi-model}.
This definition of quasi-local Pauli models is widely used in the literature~\cite{van2023probabilistic,wagner2023learning,chen2026efficient} under the name of sparse Pauli-Lindblad models~\eqref{eq:paulilind} or inclusive Pauli channels, while there also exist alternative inequivalent definitions~\cite{flammia2020efficient,rouze2023efficient}.

One major advantage of our definition is that the learnable and gauge parameters can be exactly characterized as a linear space over the noise parameters $\bm\tau$ due to the linear relation between $\bm\tau$ and $\bm x$~\cite{chen2026efficient}.
Specifically, as proven in Ref.~\cite{chen2026efficient}, for all the quasi-local Pauli noise models considered in this work, the gauge parameters can be completely described by $n$ single-qubit depolarizing channels, leading to a reduction in the number of gauge parameters from $2^n-1$ to $n$.
To learn such a quasi-local Pauli noise model up to gauge parameters, we will similarly construct a linear system of equations $\bm b = F\bm x$ (or $\bm b = F''\bm \tau$) such that the design matrix $F$ (or $F''$) reaches the maximal rank determined by the number of gauge parameters~(See Sec.~\ref{supp:constructing_design_matrix} for more details).

\section{THEORY RESULTS}

\subsection{Self-consistent error mitigation}\label{sec:theory_validation}
\begin{figure*} 
	\centering
	\includegraphics{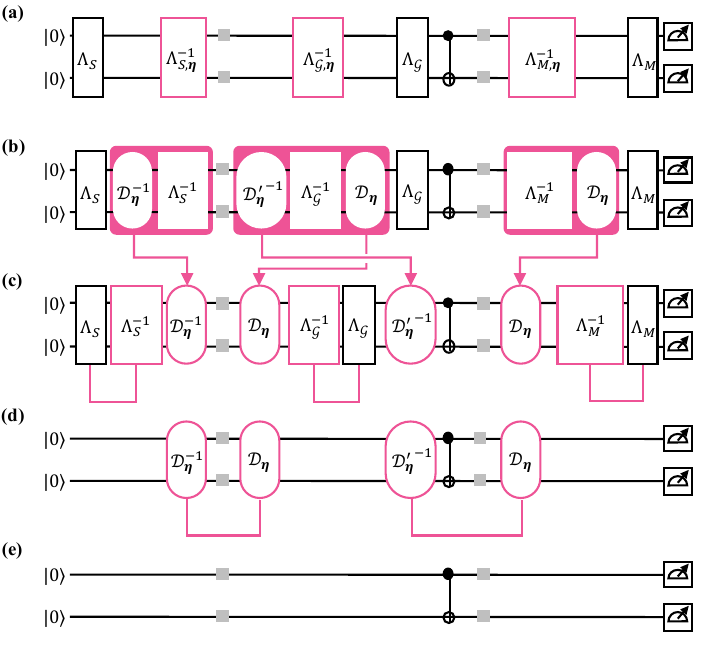}

	\caption{\textbf{Graphical proof that a complete gate set learned in a self-consistent manner can be validated using an error mitigation formalism.}
	(\textbf{a}) Using the probabilistic error cancellation (PEC) framework, the quasi-inverse channels $\Lambda_{S/\mathcal{G}/M,\bm \eta}^{-1}$ can be applied at the expense of a sampling overhead which increases exponentially with the amount of noise in the constituent noise channels. Unlike previous formulations, we apply this inverse channel in a self-consistent manner where the entire gate set shares the same set of gauge parameters $\bm \eta$. For the sake of clarity, we chose a controlled-NOT gate as the two-qubit gate. This same proof applies to any other type of two-qubit gate, and to arbitrary numbers of qubits.
	(\textbf{b}) Substitution of the gauge $\mathcal{D}_{\bm\eta}$ and Pauli noise channels $\Lambda_{S/\mathcal{G}/M}^{-1}$ from Fig.~\ref{fig:1}b yields the operations in the red boxes.
	(\textbf{c}) Reordering the gauge channels (pink arrows) leads to cancellations, or compositions of identity channels (pink brackets). Note that Pauli channels commute with each other.
	(\textbf{d}) The gauge channels and their inverses also compose to identity channels (pink brackets). Note that the generalized depolarizing channels commute with any single-qubit gates (gray squares). Also recall $\mathcal{D}_{\bm\eta}^{'}$ is defined to be the gauge channel conjugated by the entangling Clifford gate ($\mathcal{G} \circ \mathcal{D}_{\bm\eta}^{'} = \mathcal{D}_{\bm\eta} \circ \mathcal{G}$).
	(\textbf{e}) Finally, the resulting, mitigated circuit shows a noise-free operation of a controlled-NOT gate on two qubits.}
	\label{fig:4} 
\end{figure*}
A major motivation for learning quantum noise is to improve the performance of quantum computations. Quantum error mitigation (QEM) is one approach for reducing the bias in noisy quantum computations by utilizing information about the learned noise. As we discussed above, there  exists a fundamental ambiguity in quantum noise learning on account of gauge degrees of freedom, leading to previous challenges for error mitigation~\cite{chen2026efficient}. Here, we will discuss how this limitation impacted existing QEM protocols, and how we overcome the challenge by introducing a self-consistent QEM framework.

In this work, we focus on probabilistic error
cancellation (PEC), one of the few QEM protocols
with provable guarantees for achieving bias-free
estimates of expectation values given sufficient
samples and accurate noise
characterization~\cite{cai2023quantum,temme2017error,
li2017efficient}. In particular, we study PEC
protocols based on Pauli noise
models~\cite{van2023probabilistic,ferracin2024efficiently}.
Because PEC is provably unbiased under an accurate
noise model, it provides the cleanest setting to
isolate and quantify the effect of noise model
inconsistencies. However, our self-consistent noise
learning approach is not specific to PEC and applies
to any error mitigation method that relies on a
learned noise model, such as zero-noise
extrapolation
(ZNE)~\cite{kim2023evidence,haghshenas2025digitalquantummagnetismfrontier}
or tensor-network error mitigation
(TEM)~\cite{filippov2023scalable,fischer2024dynamical}.
The extension to ZNE is discussed in
Appendix~\ref{sm:ZNE}.

Let us briefly review how PEC works: consider the task of expectation value estimation for an observable on the output state of a quantum circuit, which is a natural task in, e.g., Hamiltonian simulation. If one runs the circuit on real quantum hardware, noise would corrupt the expectation value. To retrieve the noiseless value, a na\"ive approach would be to cancel out all the noise channels $\Lambda$ by implementing their inverse map $\Lambda^{-1}$. The challenge is that $\Lambda^{-1}$ is generally not completely-positive, thus not a physically realizable quantum channel. Nevertheless, when $\Lambda$ is a Pauli channel, its inverse can be formally written as $\Lambda^{-1}(\cdot)=\sum_ap_a^{\star}P_a(\cdot)P_a$ with $\sum_ap^{\star}_a=1$ but $p^{\star}_a$ can be negative.
To implement $\Lambda^{-1}$ in expectation, one can sample and apply a Pauli gate $P_a$ according to the following distribution,
\begin{equation}\label{eq:def_gamma}
    q_a = \frac{|p_a^\star|}{\gamma},\quad\text{where}~\gamma=\sum_a|p_a^\star|.
\end{equation}
where a factor of $\gamma\cdot\mr{sign}(p_a^\star)$ is then multiplied with the experimentally measured estimator, resulting in the cancellation of the noise channel $\Lambda$ in the expectation value.  Applying this procedure to cancel every noise channel in the circuit, the resulting estimation is an unbiased estimator for the noiseless expectation value. 
While this procedure works for \textit{any} quantum circuit, the trade-off is that an additional sampling overhead of $\prod_j\gamma_j^2$ is needed where $\gamma_j$ corresponds to the $j$-th noise channel, due to the multiplied pre-factors. 

Computing $p^\star$ can be computationally challenging in general. 
For quasi-local Pauli channels as in Eq.~\eqref{eq:quasi_local}, an alternative approach is given in Ref.~\cite{van2023probabilistic}.
First note that the inverse of $\Lambda$ can be written as
\begin{equation}
    \Lambda^{-1}=\bigcirc_{a\in\mc K}\left(\frac{-\omega_a}{1-2\omega_a}P_a(\cdot)P_a+\frac{1-\omega_a}{1-2\omega_a}(\cdot)\right).
\end{equation}
Then, we can simply invert each factor of $\Lambda^{-1}$. For $\omega_a\le0$, the factor is a proper Pauli channel that can be implemented without any overhead. For $0<\omega_a<1/2$, the factor can be implemented in expectation with an overhead $\gamma_a=1/(1-2\omega_a)=\exp(\tau_a)$. Multiplying the overhead from each factor yields,
\begin{equation}\label{eq:localoverhead}
    \gamma=\exp(\sum_{a\in \mathcal{K}}\max(0,\tau_a)),
\end{equation}
which we refer to as the overhead associated with the quasi-local Pauli channel $\Lambda$.

The above PEC protocol requires full knowledge of the noisy gate set so as to implement the inverse noise channels. 
However, there generically exists gauge ambiguity in learning the noise parameters, hindering a direct application of PEC.
In recent literature of error mitigation with Pauli noise models~\cite{van2023probabilistic,kim2023evidence,ferracin2024efficiently}, the issue of gauge ambiguity is circumvented by imposing the ``symmetry assumption''. As an illustration of this assumption, we consider a Clifford gate $\mc G$ that satisfies $\mc G^2 = I$ where the gate can be, for example, a CNOT gate. For any $P\in\Pn$, $Q = \mc G(P)$, whenever $P\neq Q$ up to a sign, the symmetry assumption imposes $\lambda^{\mc G}_P=\lambda^{\mc G}_Q$ (e.g. $\lambda^{\text{CNOT}}_{XI}=\lambda^{\text{CNOT}}_{XX}$). This ensures every $\lambda_P^{\mc G}$ can be uniquely determined by cycle benchmarking~\cite{flammia2020efficient,erhard2019characterizing,carignan2023error}. Furthermore, the state-preparation noise is assumed to be noiseless to determine and mitigate measurement noise~\cite{van2022model,chen2021robust}. 
We later show that these assumptions are not only unnecessary, but also lead to \textit{inconsistent} characterization of the gate set, which results in biased estimates of expectation values in applications such as QEM.

Here, we propose a self-consistent PEC protocol that properly considers the gauge parameters, thus disambiguating Pauli noise in quantum computers. Our protocol builds on the gate set Pauli noise learning framework~\cite{chen2026efficient} discussed in the last section, which enables learning a set of noise channels $\bm\Lambda_{\bm\eta}=\{\Lambda^S_{\bm\eta}, \Lambda^M_{\bm\eta},\{\Lambda_{\bm\eta}^{\mc G}\}\}$ that are gauge-equivalent to the true noise channels $\bm\Lambda=\{\Lambda^S, \Lambda^M,\{\Lambda^{\mc G}\}\}$, meaning that the two noisy gate sets have exactly the same behavior in any experiments. Thus, without ever fixing the gauge parameters, the learned $\bm\Lambda_{\bm\eta}$ contains as much information as the true noisy gate set, which can be applied to PEC. We formalize our claim in the following theorem.
\begin{theorem}[Self-consistent PEC]\label{th:main}
    Let $\bm\Lambda_{\bm\eta}=\{\Lambda^S_{\bm\eta}, \Lambda^M_{\bm\eta},\{\Lambda_{\bm\eta}^{\mc G}\}\}$ be a collection of Pauli noise channels that are gauge-equivalent to the true noise channels.
    By applying PEC as if $\bm\Lambda_{\bm\eta}$ is the ground truth, one can obtain unbiased estimators for any circuits and observables.
\end{theorem}
The formal statement and proof for Theorem~\ref{th:main} is given in Appendix~\ref{sm:gcqem_local}. An illustrative proof is provided for a two-qubit system in Fig.~\ref{fig:4}. Note that Theorem~\ref{th:main} holds even for quasi-local Pauli noise models, which is crucial for scaling up to large systems.

In light of Theorem~\ref{th:main}, it is straightforward to explain how the assumptions of symmetric gates and perfect initialization lead to inconsistency. Basically, the assumptions result in fixing each Pauli channel individually, leading to a model in the form of $\{\Lambda^S_{\bm\eta_S},\Lambda_{\bm\eta_M}^M,\{\Lambda_{\bm\eta_{\mc G}}^{\mc G}\}\}$, where an inconsistent choice of gauge parameters -- $\bm\eta_S, \bm\eta_M, \bm\eta_{\mc G}$ -- are applied for each component of the gate set. To see this in the context of Fig.~\ref{fig:4}, the residual generalized depolarizing channels remaining after mitigation would result in a biased estimation. 

We remark that the idea of combining gate set tomography (GST)~\cite{nielsen2021gate,blume2013robust} with QEM to address gauge ambiguity has been discussed in the literature~\cite{endo2018practical}. However, due to the extreme complexity and resource cost of GST, it is unclear how to apply such protocols beyond a few qubits. 
Instead, our method builds on the recently proposed gate set Pauli noise learning framework~\cite{chen2023learnability,chen2026efficient}, which enables explicit and efficient parameterization of all learnable and gauge parameters under a practical quasi-local noise assumption. 
To our knowledge, this is the first experimental demonstration of self-consistent QEM, with comparable scalability as state-of-the-art QEM protocols~\cite{van2023probabilistic,kim2023evidence,ferracin2024efficiently}.

Finally, recall PEC requires a sampling overhead $\prod_j\gamma_j^2$. Theorem~\ref{th:main} suggests that, by assuming the true noise model to be any of the gauge-equivalent models $\bm\Lambda_{\bm\eta}$, parameterized by the gauge parameters $\bm\eta$, PEC yields unbiased estimators.
Interestingly, while different $\bm\Lambda_{\bm\eta}$ all yield the same observable outcomes, different gauge parameters do not give us the same sampling overhead.
This motivates us to conduct gauge optimization -- searching for $\hat {\bm\eta^*}$ that minimizes the PEC overhead. 
The reader may wonder whether the difference in sampling overhead can be used to determine the gauge parameters $\bm\eta$. This is not possible, as the overhead merely depends on what we infer the noise parameters to be, but not what the true parameters actually are.  
In this work, we unify the gate set learning with shot noise and gauge optimization as a single convex optimization problem, which is efficiently solvable and can drastically reduce the PEC sampling overhead. 

We provide an experimental demonstration of mitigation overhead reduction through gauge optimization in Sec.~\ref{sec:gauge-optimized_PEC}.

\subsection{Efficient gauge optimization}\label{sec:gauge_opt}

Beyond addressing potential sources of bias in mitigated expectation values, we now show how the self-consistent learning approach can also be used to reduce the sampling complexity needed to successfully perform error mitigation for \textit{any} quantum circuit. 
Recall that we have $\bm b=F\bm x$; however, in this case we prefer the noise parameters in the basis of the gauge parameters $\bm r$ because it is polynomial in size for quasi-local models. The conversion from $\bm x$ to $\bm r$ is discussed in Sec.~\ref{supp:quasi-model}, which allows us to rewrite the design matrix as $\bm b = F' \bm r$.
Suppose we have a design matrix $F'$ and estimation of $\hat{\bm b}$  from experiments.
A na\"{i}ve approach to obtaining $\bm r$ and minimizing the sampling overhead needed for QEM involves first performing a pseudo-inverse of the design matrix $\hat{\bm{r}}_0={F'}^+ \hat{\bm{b}}$ (which fixes the residual error $\epsilon = \left\lVert {F'} \hat{\bm{r}}_0 - \hat{\bm{b}} \right\rVert_2$), followed directly by a second optimization step over gauge parameters $\bm\eta$ on the overhead $\gamma = \text{exp}(\sum_{a,\text{layer}} \text{max}(\tau_a^{\text{layer}},0))$ where the sum is performed over all quasi-local generators $\tau_a$ for all layers of gates, see Eq.~\eqref{eq:localoverhead}. 
However, such an approach unnecessarily restricts the gauge optimization procedure without taking into consideration that the residual errors can vary depending on the statistical fluctuations of the observed outcomes.

Rather, we introduce a \textit{one-step} optimization strategy where the possible $\bm{r}$ parameters are searched in a self-consistent manner subject to a constrained residual error $\epsilon$ chosen \textit{a priori}. 
That is, we solve the following optimization problem:
\begin{equation}
	\begin{aligned}
	&\min_{\hat{\bm{r}}}  
	\left\{
		\sum_{a\in\mc K,~\text{layer}}
		\text{max} 
		\left(\hat\tau_a^{\text{layer}}, 
			0
		\right) 
	\right\}
	\\
	&~~\text{s.t.}~  \left\lVert {F'} \hat{\bm{r}} - \hat{\bm{b}} \right\rVert_2 \le \epsilon.
	\end{aligned}
	\label{eq:gauge_opt_1step} %
\end{equation}
where the parameters $\hat{\tau}_a$ depend on $\hat{\bm r}$ following Eq.~\eqref{eq:model_to_lind} and Eq.~\eqref{eqn:mobius}. 
In Appendix~\ref{sm:convex}, we show the above gauge optimization problem can be rewritten as a second-order cone program (SOCP), thus provably solvable in polynomial time with $n$ using standard convex optimization solvers~\cite{cvxpy,lobo1998applications}, as long as the quasi-local model we use contains only polynomially many parameters.

\section{EXPERIMENTAL RESULTS}

\subsection{Restricted experiment on two qubits}
\label{sec:2qubit}

In the following we report a series of experiments that demonstrate the importance of self-consistent noise learning for error mitigation with increasing complexity of the noise models. 
In all of these experiments, we learned two noise models. The first model represents the previous state-of-the-art~\cite{van2023probabilistic, kim2023evidence, van2022model} which imposes the symmetry assumption between conjugate Pauli eigenvalues and assumes ideal state preparation for readout error mitigation. We refer to this as the ``inconsistent'' noise model. 
The second model learns all noise channels in a self-consistent way and is referred to as the ``consistent'' noise model. 
For all experiments, we compare the performance of both models in predicting noisy expectation values in the corresponding circuits, which is numerically tractable due to the Clifford nature of the circuits. 

As a first step, we examined the impact of self-consistent learning for a gate set on two qubits. Recall that methods which use a framework lacking this consistency may work on some experiments but fail on others. To highlight the difference between such learning frameworks, we used a 27-qubit device named \textit{ibm\_auckland} with calibrated CNOT gates as the basis two-qubit gates. In this case, our full gate set was composed of initialization to $|0\rangle^{\otimes2}$, a single CNOT gate, any single-qubit gates (with negligible noise), and computational basis measurements. A rigorous way to represent all possible experimental outcomes on this two-qubit system is captured in the pattern transfer graph described earlier (Fig.~\ref{fig:2}).

\begin{figure*} 
	\centering
	\includegraphics[width=0.75\textwidth]{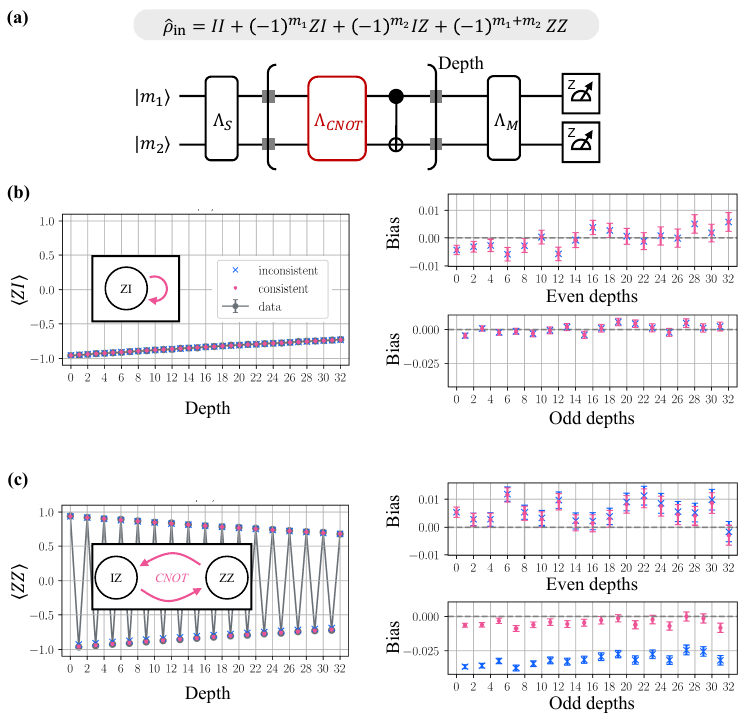} 
	\caption{\textbf{Experimental learning and mitigation of a restricted set of errors on two qubits.}
	(\textbf{a}) Circuit used for both learning and targeted mitigation, except an initial state with $m_1=m_2=0$ is used for the learning while $m_1=m_2=1$ is used for the target circuit. This restricts learning to three Pauli terms: $IZ$, $ZZ$ and $ZI$. Note that the target circuit is prepared in $|11\rangle$, the $-IZ$ and $+ZZ$ eigenstates, which differs from the learning circuit which is prepared in $|00\rangle$, the $+IZ$ and $+ZZ$ eigenstates. The ``inconsistent'' noise model only requires learning with circuits using even depths (0, 2, 4,.. 32), while the ``consistent'' learning model requires one additional depth-1 experiment. 
	(\textbf{b, c}) For the non-degenerate (b) and degenerate (c) cycles, the experimental data (gray) from the target circuit is shown alongside the predicted outcomes using the inconsistent (blue) and consistent (pink) noise models. To the right of (b), both the even and odd depths show no difference in predicted outcomes. However, to the right of (c), the even depth shows no difference while the odd depths show a difference of 3.2$\pm$0.4\% (blue, inconsistent) compared to a 0.5$\pm$0.3\% (pink, consistent) bias in the outcomes.
		}
	\label{fig:5}
\end{figure*}

Rather than examining all the cycles in the pattern transfer graph, we prepared and measured experiments in the $Z$-basis exclusively, and thus we needed to focus on learning only two specific cycles: the $ZI\circlearrowleft$ and $IZ \leftrightarrow ZZ$. While one of these cycles involves only a single node (``10''), the other involves two nodes (``01'' and ``11''). 
The latter is referred to as a degenerate cycle or a conjugate pair as they contain two eigenvalues that cannot be separately determined~\cite{chen2023learnability,ferracin2024efficiently,van2023probabilistic}.

A convenient consequence for focusing on the $Z$-only eigenvalues means the preparation and measurement bases can be entirely in the computational states (i.e. $|0\rangle$ or $|1\rangle$). 
Thus, to learn cycles restricted to a certain type - in this case the Z-only observables - we only needed to prepare the $|00\rangle$ initial state for circuits with increasing repetitions of the CNOT gate from depths 0 to 32. 
Unlike previous noise learning approaches which only utilized circuits with even numbers of CNOT gates, here we also introduced a learning circuit with just a single, or depth-1, application of CNOT. 
For larger experiments discussed later, more preparation and measurement bases for the depth-1 experiments need to be incorporated to learn all possible learnable parameters described under Eq.~\eqref{eq:design_matrix}. 
Intuitively, the reason depth-1 experiments are needed for self-consistent learning is to account for the degeneracy between the conjugate Paulis -- in this case between $IZ$ and $ZZ$. 
Thus for the same set of experiments, we were able to learn two noise models for the eigenvalues $\lambda_{ZI}$, $\lambda_{IZ}$, and $\lambda_{ZZ}$: one that is \textit{self-consistent} and another \textit{inconsistent} which assumes that any conjugate Paulis are symmetric and does not incorporate depth-1 results. 

To compare the validity of the two noise learning approaches, we then separately performed so-called ``target'' experiments where the observables $\langle ZI\rangle$ and $\langle ZZ\rangle$ were measured after initialization into the $|11\rangle$ state; we compared the outcomes against those predicted from the learned noise models using only the $|00\rangle$ initial state. Ideally, the experimentally measured outcomes should agree with those predicted from the noise models, and thus their division should yield an ideal value of $1$ - known in this case because these are Clifford operations on an initial stabilizer state. The ratio of the two values - measured to the predicted - informs us how much mitigation bias persists based on the different noise models. 

We show the experimental outcomes, along with those predicted by the two different noise learning procedures, in Fig.~\ref{fig:5}b, c. 
As expected, the non-degenerate cycle involving the $\langle ZI\rangle$ observable exhibited no difference in bias between the experimentally measured and the predicted outcomes from two noise learning approaches at even or odd depths. The reason we separated out the even from odd depths is because the ``inconsistent'' noise model, in this special case, is unambiguous at predicting outcomes of circuits for depth-even applications of CNOTs. 

In contrast, the degenerate cycle $IZ \leftrightarrow ZZ$ showed no mitigation bias when an even number of CNOT gates were applied, as expected based on the fact that the ``inconsistent'' model accurately captures the product of the conjugate Pauli eigenvalues. 
However, for odd-depths applications of the CNOT gate, we find that the self-consistent learning protocol reduces the predicted bias from 3.2$\pm$0.4\% down to 0.5$\pm$0.3\% when averaged over all 16, odd layer depths of the target circuit up to depth 31. 
This statistically significant improvement in predicted outcomes using the self-consistent learning approach was reproduced across a total of six qubit pairs on the same device (Data in Appendix~\ref{sm:more_pair}). 
Despite the restricted nature of this experiment on only two qubits, the widespread improvement in mitigation bias with self-consistent learning motivated the question of how much this bias can be improved for circuits with more qubits.

\subsection{Restricted experiment on entangled states}\label{sec:GHZ}
\begin{figure*} 
	\centering
	\includegraphics[width=1\textwidth]{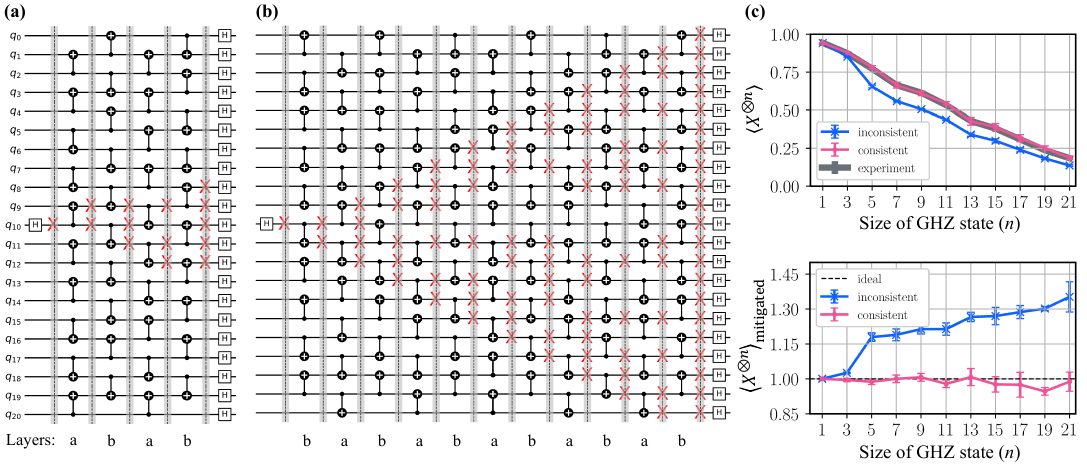}
	\caption{\textbf{Self-consistent experimental learning of non-local errors and mitigation of a GHZ state on up to 21 qubits.}
	(\textbf{a}) Quantum circuit used to prepare a 5-qubit GHZ state using two template layers: `a' and `b', where layers from the same template do not necessarily have the same direction of controlled-not gates. The root qubit, where the a Hadamard (H) gate is applied before the first layer, can be described by a weight-1 stabilizer $X$ (red), which is shown to grow to weight-5 after four total layers.
	(\textbf{b}) Quantum circuit used to prepare a 21-qubit GHZ state using the same two template layers, except the controlled-not gate directions again do not match with the circuit used to prepare the 5-qubit GHZ state in (a). The growth of the $\langle X^{\otimes n} \rangle$ stabilizer for the GHZ state reaches full-weight by the end of the circuit as indicated by the column of red $X$ labels immediately before the final layer of Hadamard gates.
	(\textbf{c}) Experimental outcomes (black line) plotted against predicted expectation values using self-consistent (pink circles) and inconsistent (blue crosses) noise models.
	(\textbf{d}) Mitigated values of the full-weight, $\langle X^{\otimes n} \rangle$ stabilizers of the GHZ state for circuits up to $n=21$ where the self-consistent noise model (pink circles) shows strong agreement with the expected value of $\langle X^{\otimes n}\rangle=1$ while the symmetric noise model (blue crosses) show increasing bias with the size of the GHZ state. The error bars for both (c) and (d) were the result of averaging over seven separate experiments~(Appendix~\ref{sm:GHZexpt} for more details).
		}
	\label{fig:6}
\end{figure*}
Next, we examined the impact of self-consistent noise learning for a target circuit with not only many more qubits, but also an observable that depends mostly on individual fidelities from degenerate cycles. 
We identified an observable such that the observed bias should increase with system size when compared against the ``inconsistent'' noise learning approach. Meanwhile we expect the ``consistent'' noise learning approach, which captures the degenerate Pauli pairs, to remain unbiased no matter the system size. 

For this task, we identified the highly entangled GHZ state as the ideal target state.
The GHZ state on $n$ qubits is a  stabilizer state that is specified by being the simultaneous +1 eigenstate of the generators, a set which includes a full-weight term $\langle X^{\otimes n}\rangle$ and $n-1$ weight-2 terms $\langle Z_i Z_{i+1}\rangle$ for $i\in [n-1]$. 
However, rather than relying on well-known preparation circuits which require learning $O(n)$ unique layers of entangling gates, we prepared the state using only two unique dense layers of simultaneously applied CNOT gates. 
Specifically, we first chose a fixed set of 21 qubits. We then  employed a SAT-solver to prepare GHZ states on $n=1,3,\dots,21$-qubit subsets of those 21 qubits using  the two unique layers of entangling gates covering all 21 qubits~\cite{gavrielov2024linear,yoshioka2024diagonalization}. A graphical way to check how our procedure works can be seen in Fig.~\ref{fig:6}a and b for $n=5$ and $n=21$ GHZ states, respectively. In those figures, we track the creation of the $\langle X^{\otimes n}\rangle$ stabilizer, and show that it grows monotonically with densely populated CNOT layers of gates.

We call the two alternating template layers `a' and `b'. 
We fix the CNOT directions for each template layer, and use interleaving single-qubit gates to effectively change the CNOT directions to arrive at the circuits in Fig.~\ref{fig:6}.
For this experiment where we are only examining the impact of self-consistent learning on the $\langle X^{\otimes n}\rangle$ observable of the target GHZ state, we only learned the Pauli eigenvalues which contribute to the construction of the final observable. 
This was only possible because our target circuit is a Clifford circuit, which means we were able to classically back-propagate the observable through the entire circuit and identify those Pauli eigenvalues needed from each instance of the two template layers. 
In this sense, this was a \textit{restricted} noise model because we did not learn the full Pauli noise channel for both layers, but allowed for the possibility of \textit{nonlocal} noise by not imposing any locality constraints; in other words, we allowed the number of gauge parameters to remain in the most general form with $2^n-1$ terms.
In Sec.~\ref{supp:constructing_design_matrix}, we include an example for learning the noise of template `a' used in preparing the $n=21$ GHZ state.

Unlike the previous section, the experiments here and in the subsequent sections were conducted using a larger, 127-qubit device named \textit{ibm\_strasbourg}. 
Similar to how we compared the learned noise models against the target circuit earlier, we again compare the predicted outcomes for the $\langle X^{\otimes n}\rangle$ observable for GHZ states with increasing sizes up to 21 qubits against the experimentally measured values (Fig.~\ref{fig:6}c, d). 
The Clifford nature of the circuit allows us to predict the resulting bias of a hypothetical mitigation experiment with PEC by dividing noisy expectation values by the values predicted by the respective models. We refer to these as ``mitigated'' values. 
Indeed, for GHZ states up to 21 qubits, we observed an increasing bias using the inconsistent noise model reaching 35.2\%$\pm$6.5\%, while we observed statistically insignificant -1.2\%$\pm$4.1\% biases using the self-consistent noise model for the largest depths. 

\subsection{Scalable learning for general, \\quasi-local noise}\label{sec:ring}
\begin{figure*} 
	\centering
	\includegraphics[width=1\textwidth]{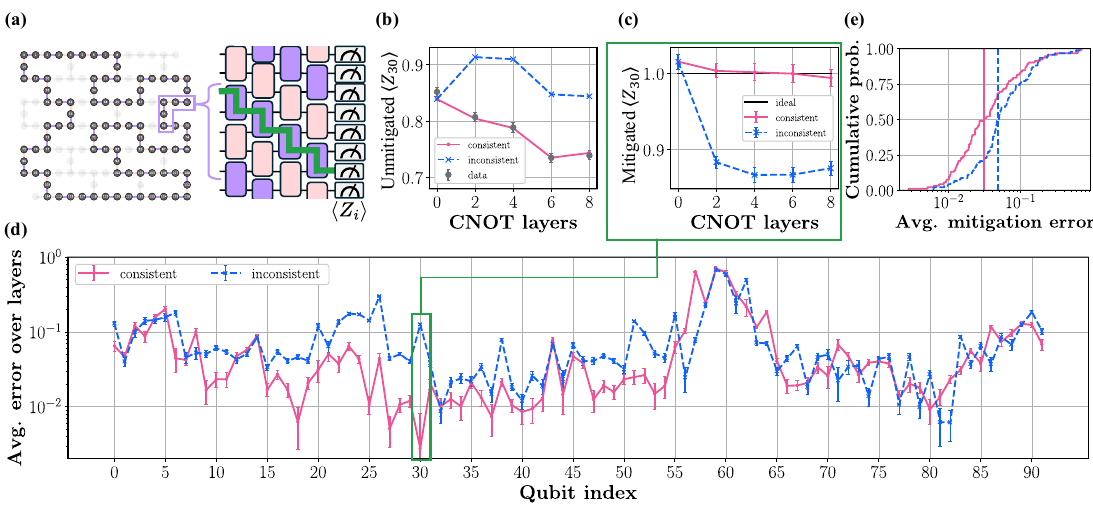} 
	\caption{\textbf{Scalable (quasi-local) self-consistent noise learning and mitigation of weight-1 observables on a ring of 92 qubits.}
	(\textbf{a}) Closed loop of 92 qubits on a 127-qubit device, \textit{ibm\_strasbourg}. Boxed section of ring shows a cross section of the 92 qubit, depth-8 quantum circuit designed specifically to propagate all 92, weight-1 stabilizers in a ``staircase'' (green) fashion such that the initial and final qubit support of the stabilizer falls on a different qubit.
	(\textbf{b}) Experimentally measured (filled gray circle) expectation values versus number of circuit layers compared against the self-consistent (solid red) and symmetric noise predictions (dashed blue).
	(\textbf{c}) Predicted values divided by the measured values yield a mitigated value for the data set in (b), and boxed in (a).
	(\textbf{d}) Average mitigation error up to four circuit layers on all 92 qubits calculated in the same manner as for a single qubit as seen in (c).
    Error bars depict one standard deviation of the shot noise on the unmitigated data. 
	(\textbf{e}) Cumulative distribution of mitigation errors (d) between the experimentally measured and predicted outcomes. The median mitigation error, denoted by vertical lines, shows a reduction from 4.9\% (dashed blue) to 3.1\% (solid red) bias.
		}
	\label{fig:7} 
\end{figure*}
Earlier we focused on restricted models with learning circuits that are straightforward to construct; now we will focus on complex learning circuits with minimal assumptions needed for constructing the full noise models. 
That is, the previous two experiments used some knowledge of the target circuit to inform the design of the learning experiments, while in this section we will discuss how to conduct complete self-consistent noise learning when given only the quasi-local noise assumption based on qubit connectivity and the template gate layers being learned. 
By providing an explicit construction for the gate layers in the gate set and a noise ansatz (e.g. 1- or 2-local), we used the formalism shown in Fig.~\ref{fig:3} and in Ref.~\cite{chen2026efficient} to construct the preparation and measurement bases needed for the learning circuits such that a self-consistent noise model could be inferred. 
Since the noise is assumed to be quasi-local, the number of parameters is no longer exponential but in fact only linear, and thus can be efficiently learned.

For $n$ qubits on a ring, we consider a gate set consisting of two gate layers $G_{\rm{even}}= {\rm{CNOT}}_{1,2}\otimes{\rm{CNOT}}_{3,4}\otimes\dots\otimes{\rm{CNOT}}_{n-1,n}$ and $G_{\rm{odd}}= {\rm{CNOT}}_{2,3}\otimes{\rm{CNOT}}_{4,5}\otimes\dots\otimes{\rm{CNOT}}_{n,1}$. The noise on each layer and on SPAM operations is assumed to be \textit{quasi}-local, i.e., it factors into a composition of channels that act only on nearest neighbor qubits. As shown in Ref.~\cite{chen2026efficient}, this model has $28n$ parameters with a fully local gauge. That is, there are $27n$ learnable parameters and $n$ gauge parameters corresponding to $n$ single-qubit depolarization maps. 

\begin{figure*} 
	\centering
	\includegraphics[width=1\textwidth]{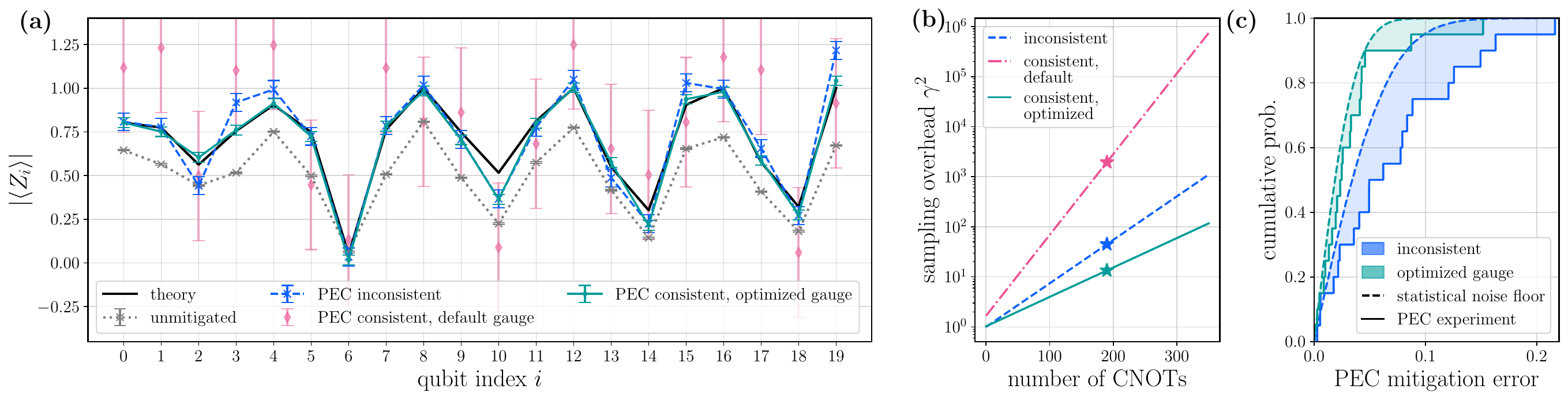} 
	\caption{\textbf{Probabilistic error cancellation with self-consistent noise models and gauge optimization.}\\
	(\textbf{a}) PEC results for single-qubit $Z$ observables in a 20-qubit non-Clifford benchmark circuit with 20 CNOT layers. Error bars show the sample standard deviation of the PEC estimators. 
    (\textbf{b}) The PEC sampling overhead $\gamma^2$ as a function of the number of CNOT gates for the characterized noise models, where the star corresponds to the circuit from panel a.      
    (\textbf{c}) 
    Cumulative distribution of the absolute remaining error of the PEC estimators for the inconsistent noise model and the gauge-optimized noise model (solid lines). Dashed lines show the statistical noise floor of unbiased PEC from repeated numerical simulations with the same number of shots and $\gamma$-values.
    Shaded regions reflect the systematic error beyond what is expected from shot noise alone. }
	\label{fig:PEC_results}
\end{figure*}

Due to the locality of the noise model, only local expectation values are needed to learn the model parameters. This allows many parameters to be estimated in parallel. As a result, the number of measurement settings needed to learn the complete noise model remains constant and does not scale with system size. For details of the experiment and specific measurement settings, see Appendix~\ref{sm:ring_experiments} and Table~\ref{tab:ring_bases}.  

We applied the learned noise model to a target circuit where we measure local $Z$ observables for every qubit on a closed ring. 
As before, the circuit consists of two layers of CNOT gates between even or odd neighboring qubit pairs.
Specifically, this is a Clifford circuit designed in a way such that the 92 observables each depend only on weight-1 and weight-2 Pauli eigenvalues which all originate from different degenerate cycles and are thus sensitive to the symmetry assumptions imposed by the inconsistent model. 
In total, the Pauli eigenvalues probed by the observables cover \text{all} degenerate Pauli eigenvalues of the participating CNOTs (See Appendix~\ref{sm:ring_experiments} for details). 
We used a ring of 92 of the 127 qubits available on \textit{ibm\_strasbourg} shown in Fig.~\ref{fig:7}a. 
With every layer of two-qubit blocks, each weight-1 eigenstate is propagated to another weight-1 eigenstate shifted by one qubit index along the ring.
Note that one such block consists of two layers of parallel CNOT gates as depicted in Fig.~\ref{fig:methods_ring_circuit}b. 
We highlight one of the 92 available observables, and show how it evolved for different circuit depths in Fig.~\ref{fig:7}b. 
Then, we compare the experimental outcomes against the predicted outcomes based on the consistent and inconsistent noise models by computing the mitigated values as before.
In Fig.~\ref{fig:7}c, we show one specific example where the bias using the inconsistent model reaches 12\%$\pm$0.5\% whereas the consistent model shows no statistically significant bias of 0.3$\pm$0.5\%. Applying the same analysis as described above across all 92 qubits, we saw that the consistent noise model yielded mitigation errors at or below that predicted with the inconsistent noise model (Fig.~\ref{fig:7}d, e). 
In fact, the median mitigation error was reduced from 4.9\% to 3.1\%. The remaining bias can be largely attributed to out-of-model errors~\cite{govia2024bounding}. More specifically, there are several physically relevant error sources that fall outside the two-local Pauli-Lindblad noise ansatz used here. Non-Markovian effects, such as coupling to two-level systems (TLS)~\cite{kim2024error}, slow temporal drift of the noise parameters between learning and target circuits, or coherent leakage out of the qubit subspace, are not captured by our noise model. Similarly, while we allow arbitrary two-body interactions between nearest-neighbor qubits, three-body correlations or long-range coherent crosstalk between non-nearest-neighbor qubits lie outside the quasi-local ansatz. While some of these errors, such as long-range crosstalk, can in principle be twirled into Pauli channels, the resulting channels are not guaranteed to be representable by a quasi-local Pauli-Lindblad generator. Others, such as leakage, cannot be twirled into Pauli noise at all. In either case, such errors fall outside the model class assumed here. As analyzed in Ref.~\cite{govia2024bounding}, the resulting mitigation bias from non‑local interactions scales linearly with the number $N$ of such terms as $\sim 4N\lambda$, where $\lambda$ characterizes their strength, implying that these effects can become more pronounced at larger system sizes. In practice, temporal drifts can be mitigated by more frequent noise characterization, while non‑local crosstalk can be substantially suppressed using context‑aware dynamical decoupling and compilation techniques~\cite{seif2024suppressing}. We stress that these limitations are shared by all model‑based error mitigation approaches, including PEC, ZNE, and TEM, and are not specific to the self‑consistent framework introduced here. Rather, the present work eliminates gauge inconsistency as a dominant and previously unaccounted‑for source of systematic error, so that the remaining bias can be more cleanly attributed to these genuine physical model violations rather than to artifacts of the learning and mitigation pipeline.

\subsection{Gauged-optimized PEC}
\label{sec:gauge-optimized_PEC}

The experiments presented so far have established that inconsistently learned noise models lead to systematic biases when predicting noisy expectation values. 
In the previous sections, this effect could be cleanly identified because the employed benchmark circuits were Clifford, allowing for exact classical predictions of the noisy outcomes. 
We now go one step further and demonstrate that the same inconsistencies lead to biased \emph{mitigated} expectation values when applying PEC (see Sec.~\ref{sec:theory_validation}) to non-Clifford circuits. 
Moreover, our experiment shows that gauge optimization can substantially improve the sampling overhead of PEC.

As benchmark circuits, we considered the circuits introduced in Sec.~\ref{sec:ring}, but perturbed them away from Clifford angles as described in Appendix~\ref{sm:pec_circuits} and again focused on single-qubit $Z$ operators. 
All experiments were performed on the \textit{ibm\_fez} device at a system size of 20 qubits, for which exact classical predictions of the ideal expectation values remain tractable. 
The noise affecting these circuits was characterized using the same procedure as in Sec.~\ref{sm:PEC_shot_noise}, assuming a quasi-local sparse Pauli-Lindblad noise model.

For the self-consistently learned noise model, we performed gauge optimization following the method described in Sec.~\ref{sec:gauge_opt}. The fitting tolerance $\epsilon$ in Eq.~\eqref{eq:gauge_opt_1step} was chosen to be consistent with the shot noise present in the learning data, as detailed in Appendix~\ref{sm:PEC_shot_noise}. 
For the benchmark circuits considered here, this results in markedly different PEC overheads: the inconsistent model yields a mitigation overhead of $\gamma = 6.74$, the self-consistent model obtained via direct inversion of Eq.~\eqref{eq:design_matrix} (which we refer to as the \emph{default} gauge choice) yields $\gamma = 48.68$, while the gauge-optimized self-consistent model achieves a substantially reduced overhead of $\gamma = 3.7$.

We applied PEC using these three noise models, see Fig.~\ref{fig:PEC_results}(a) for the resulting mitigated expectation values. 
For each model, PEC was performed using $100{,}000$ randomized circuit instances, which include both quasi-probabilistic sampling of the inverse noise channels and Pauli twirling, with a single shot taken per instance. 
While state preparation and gate errors are mitigated via genuine PEC sampling, measurement errors were corrected in post-processing in the spirit of TREX: We divided the estimators by the corresponding readout fidelities (or, in the case of the inconsistent model, by the inferred SPAM fidelities) associated with the measured observables. 
To further suppress non-Markovian effects, we employ a post-selection strategy that detects leakage out of the computational subspace~\cite{jaytalk, Kim2026InPrep}. After this procedure, we retained approximately $18\,\%$ of the $100{,}000$ collected shots.

As expected, the mitigated values using the default self-consistent model exhibit very large statistical error bars due to its large associated $\gamma$ value. 
The inconsistent model, on the other hand, displays a clear residual bias, which we attribute to the same systematic modeling errors identified in the previous sections. 
In contrast, the gauge-optimized self-consistent model performs best overall, simultaneously exhibiting lower bias and reduced variance. 
The difference in PEC sampling overhead for the three models is further illustrated in Fig.~\ref{fig:PEC_results}(b). We note the offset of the curves corresponding to the self-consistent models, which arises from the additional overhead associated with mitigating state preparation errors—an overhead that is absent by construction in the inconsistent model.

Quantitatively, PEC using the inconsistent model reduces the median error from $21.8\%$ (unmitigated) down to $5.6\%$. This is further improved to a median error of $2.4\%$ when using the gauge-optimized self-consistent model; the full error distributions are shown in Fig.~\ref{fig:PEC_results}(c). 
In the same figure, we compare the observed mitigation errors to the expected statistical noise floor obtained from simulations of unbiased PEC with the same number of samples and $\gamma$ values. 
We find that the remaining errors for the gauge-optimized model are largely consistent with residual shot noise, whereas the inconsistent model exhibits a pronounced deviation from this statistical noise floor, indicating the presence of additional systematic errors. 
This observation is consistent with the conclusions drawn from the Clifford experiments presented earlier in this work.
Taken together, these results provide direct experimental evidence that gauge optimization can substantially enhance the practical performance of error mitigation, both by reducing bias and by significantly lowering the associated sampling overhead.

Our experiment represents one of the most rigorous realization of PEC to date, surpassing all previous implementations in the number of sampled circuit instances. 
Moreover, previous works on noise-model-based error mitigation have implored Markovian models with non-negative generator rates~\cite{van2023probabilistic, kim2023evidence, fischer2024dynamical}. 
Our implementation generalizes this to non-Markovian noise models that include negative generator rates, as recently proposed in independent theoretical works~\cite{seif_single_2026, kattemolle2026non}.

\section{DISCUSSION}
As noise in quantum computers continuously improves, quantum error mitigation will become increasingly effective at unlocking some of the potential applications promised by fault-tolerant quantum computers~\cite{aharonov2025importanceerrormitigationquantum}. While the idea of leveraging improved noise learning procedures for error mitigation was proposed~\cite{endo2018practical}, there was no proposal, to the best of our knowledge, to demonstrate the idea in a scalable manner.
By showing that an accurate, self-consistent noise learning framework can be utilized for one of the most prominent error mitigation techniques, we have taken an important step towards realizing practically useful applications on pre-fault tolerant quantum systems. While  learning a quantum process by itself can be a candidate for quantum advantage~\cite{huang2022quantum}, for example by utilizing entanglement to obtain a substantially lower sample complexity~\cite{PhysRevA.105.032435,seif2024entanglement}, it can also be used for more precise diagnosis of the most immediate hardware or material limitations to be addressed~\cite{de2021materials}. 

Our method resolves the issues arising from treating noise in different components of an experiment inconsistently, e.g., choosing different gauges for SPAM and gate errors. While it does not uncover the true, unobservable gauge, we anticipate that combining this approach with insights from the underlying physics of the processes can lead to a more accurate characterization of the actual noise affecting operations. For example, a more detailed understanding of entangling gates~\cite{malekakhlagh2025efficient} and the different mechanisms involved in state preparation versus measurement  may help determine a more physically meaningful gauge.

Our experimental approach, while requiring a constant number of additional circuits to learn the noise, results in verifiable accuracy for both deep and large circuits similar to those used in near-term and long-term application circuits. Unlike other approaches whose formalism depends heavily on the locality of the physical noise, our experimental design can be catered to any noise ansatz based on the underlying quantum hardware. Furthermore, once the noise is accurately learned, an important application of this framework is that the associated sampling overhead needed for error mitigation can be reduced.

It would be intriguing to extend this self-consistent noise formalism to dynamic circuits where subsequent classical operations can depend on the outcomes of mid-circuit measurements~\cite{zhang2025generalized, hines2025pauli}. 
Such dynamic circuits are considered promising for preparing and simulating interesting states with significantly less circuit depth~\cite{tantivasadakarn2023hierarchy, buhrman2024state, gupta2024probabilistic}. 
For hybrid quantum-classical computations, dynamic circuits are also believed to be free of barren plateaus~\cite{deshpande2024dynamic}. 
Being able to mitigate such mid-circuit measurement errors, once accurately learned in a self-consistent manner, can open up new avenues for quantum error mitigation~\cite{chen2025nishimori, baumer2024efficient}. 
More accurate noise models of such non-unitary operations are also essential for optimizing the performance of decoders needed to \textit{actively} correct errors in large-scale, fault-tolerant quantum computers~\cite{PhysRevLett.128.110504,bausch2024learning}.

\begin{acknowledgments}
We are grateful for helpful discussions with Zlatko Minev, Maika Takita, Abhinav Kandala, Alexander~Ivrii, David Layden, Ian Hincks, Sam Ferracin, James Raftery, Blake Johnson, Steve Flammia, Zhihan Zhang, Yunchao Liu, Adrian Chapman, and Sisi Zhou.
This work has been supported by the IBM-UChicago Quantum Collaboration, under agreement number MAS000364, with access to the fleet of IBM Quantum computers.
S.C. and L.J. acknowledge support from the ARO(W911NF-23-1-0077), ARO MURI (W911NF-21-1-0325), AFOSR MURI (FA9550-21-1-0209, FA9550-23-1-0338), NSF (OMA-2137642, OSI-2326767, CCF-2312755, OSI-2426975), and the Packard Foundation (2020-71479).
L.E.F. acknowledges funding from the European Union’s Horizon 2020 research and innovation program under the Marie Sk\l{}odowska-Curie grant agreement No.~955479 (MOQS – Molecular Quantum Simulations). 
\end{acknowledgments}

\clearpage 

\bibliography{gauge} 
\newpage

\appendix

\widetext

\section{Proof for self-consistent error mitigation}
\label{sm:gcqem_local}
In this section, we give a rigorous proof for Theorem~1. For this purpose, we will first review the standard PEC procedure, prove its correctness, and then generalizes to self-consistent PEC.

Let us first specify the model assumptions.
For an $n$-qubit system, we consider the following set of operations and their noisy implementation.
\begin{enumerate}
    \item Initialization: $\ketbra{0}{0}\mapsto \tilde\rho_0 = \Lambda_S(\ketbra{0}{0})$.
    \item Computational-basis measurement: $\{\ketbra{b}{b}\}_{b\in\{0,1\}^n}\mapsto\{\tilde E_b = \Lambda_M(\ketbra{b}{b})\}_{b\in\{0,1\}^n}$.
    \item Layer of single-qubit unitary: $\mc U=\otimes_{k=1}^n\,\mc U_k$, implemented without noise.
    \item Layer of multi-qubit Clifford: $\mc G\mapsto\tilde{\mc G}={\mc G}\circ\Lambda_{\mc G}$, for all $\mc G$ from a finite set $\mf G$.
\end{enumerate}
Here, we further assume $\Lambda_{\mc G}$ are $\mc G$-dependent Pauli channels, and  $\Lambda_S,\Lambda_M$ are generalized depolarizing channels (i.e., $\lambda_a$ only depends on $\pt(a)$).  
We use $\bm\Lambda$ to denote the collection of all noise channels. 
Furthermore, we assume all the Pauli eigenvalues are strictly positive.
All these assumptions are experimentally justified via randomized compiling~\cite{wallman2016noise}. We also allow these Pauli channels to come from any quasi-local ansatzes, as introduced in the main text.

Though we only define the noise channel associated with the computational-basis measurement, since we assume single-qubit gates to be noiseless and $\Lambda_M$ to be invariant under single-qubit rotation, we can effectively estimate any observable $O$ up to the $\Lambda_M$, i.e., $\tilde O =\Lambda^M(O)$.

\paragraph{Standard PEC.} Suppose we want to estimate the expectation value of an observable $O$ at the output state of a quantum circuit. Denote the ideal value by
\begin{equation}
    o = \lbra O \mc G_T\mc U_T\cdots\mc G_1\mc U_1 \lket{\rho_0}.
\end{equation}
Here, $\mc U_j$'s are layers of (possibly non-Clifford) single-qubit gates, and $\mc G_j$'s are layers of multi-qubit Clifford gates from $\mf G$. Because of noise, a direct execution of the above gate sequence will instead give
\begin{equation}
\begin{aligned}
    o^{(\mr{noisy})} &= \lbra{ \widetilde O }\widetilde{\mc {G}}_T\mc U_T\cdots\widetilde{\mc G}_1\mc U_1 \lket{\widetilde{\rho}_0}\\
    &=\lbra O \Lambda_M\mc G_T\Lambda_{\mc G_T}\mc U_T\cdots\mc G_1\Lambda_{\mc G_1}\mc U_1 \Lambda_S\lket{\rho_0}.
\end{aligned}
\end{equation}
To retrieve the ideal value, a naive idea is to cancel out all noise channels $\Lambda$ by implementing $\Lambda^{-1}$. For any Pauli channel $\Lambda=\sum_b\lambda_b\lketbra{P_b}{P_b}/2^n$, its inverse is $\Lambda^{-1}=\sum_b\lambda_b^{-1}\lketbra{P_b}{P_b}/2^n$, which can be expressed as
\begin{equation}
    \Lambda^{-1}(\rho) = \sum_{a\in\Pn}p^\star_aP_a\rho P_a,\quad\mr{where}~~p^\star_a=\frac{1}{4^n}\sum_{b\in\Pn}(-1)^{\expval{a,b}}\lambda_b^{-1}.
\end{equation}
This is a Pauli diagonal map that is not necessarily completely-positive (i.e., $p_a^\star$ can be negative). Consequently, it cannot be directly implemented as a quantum channel. Instead, one can rewrite it in the following form
\begin{equation}
\begin{aligned}
    \Lambda^{-1}(\rho) &= \sum_a\left(\sum_b{|p_b^\star|}\right)\frac{\mr{sgn}_a|p_a^\star|}{\sum_b|p_b^\star|}P_a\rho P_a\\
    &= \sum_a  \gamma \mr{sgn}_a q_a P_a\rho P_a,
\end{aligned}
\end{equation}
where $\mr{sgn}_a$ is the sign of $p_a^\star$, $\gamma =\sum_b|p_b^\star|$, and $q_a = |p_a^\star|/\gamma$. Note that $\{q_a\}$ forms a probability distribution over $\Pn$. Thus, by sampling Pauli operator $P_a$ according to $\{q_a\}$ and multiplying $\gamma\mr{sgn}_a$ in classical post-processing, one can implement $\Lambda^{-1}$ in expectation. This is the core idea of PEC. 

Concretely, consider the following steps of \emph{standard PEC} (which has assumed all noise channels are known a priori):
\begin{enumerate}
    \item Randomly sample $a_0\sim\{q_{a_0}^S\}$, $a_j\sim\{q^{\mc G_j}_{a_j}\}_{j=1}^T$,  $a_{T+1}\sim\{q_{a_{T+1}}^M\}$. 
    \item Implement and measure the following expectation value
    \begin{equation}\label{eq:random_sm_eq1}
        E_{\bm a} = \lbra{ \widetilde O }\mc P_{a_{T+1}}\widetilde{\mc {G}}_T\mc P_{a_{T}}\mc U_T\cdots\widetilde{\mc G}_1\mc P_{a_{1}}\mc U_1\mc P_{a_{0}} \lket{\widetilde{\rho}_0}
    \end{equation}
    where $\mc P_a(\rho) = P_a\rho P_a$.
    \item Define the PEC estimator as
    \begin{equation}
        \hat o^{\mr{(PEC)}} = E_{\bm a}\cdot\prod_{j=0}^{T+1}\gamma_j\mr{sgn}_{a_j}.
    \end{equation}
    where $\gamma_j$ and $\mr{sgn}_{a_j}$ are with respect to the $j$th Pauli noise channel.
\end{enumerate}

The following proposition shows the correctness of PEC.
\begin{proposition}\label{prop:std-pec}
    Given that one knows $\bm\Lambda$ exactly, the standard PEC estimator $\hat o^{\mr{(PEC)}}$ is an unbiased estimator for $o$.
\end{proposition}

\noindent\textbf{Proof.}
    \begin{equation}
    \begin{aligned}
        \E {\hat o^{\mr{(PEC)}}} &= \sum_{\bm a}E_{\bm a}\cdot\prod_{j=1}^{T+1}\gamma_j\mr{sgn}_{a_j}q_{a_j}\\
        &=\sum_{\bm a}E_{\bm a}\cdot\prod_{j=1}^{T+1}p_{a_j}^\star\\
        &=\sum_{\bm a}\lbra{ \widetilde O }p^{\star,M}_{a_{T+1}}\mc P_{a_{T+1}}\widetilde{\mc {G}}_Tp^{\star,\mc G_T}_{a_{T}}\mc P_{a_{T}}\mc U_T\cdots \widetilde{\mc G}_1p^{\star,\mc G_1}_{a_{1}}\mc P_{a_{1}}\mc U_1p^{\star,S}_{a_{0}}\mc P_{a_{0}} \lket{\widetilde{\rho}_0}\\
        &= \lbra{ \widetilde O }\Lambda_M^{-1}\widetilde{\mc {G}}_T\Lambda_{\mc G_T}^{-1}\mc U_T\cdots\widetilde{\mc G}_1\Lambda_{\mc G_1}^{-1}\mc U_1\Lambda_0^{-1} \lket{\widetilde{\rho}_0}\\
        &= \lbra O \mc G_T\mc U_T\cdots\mc G_1\mc U_1 \lket{\rho_0}\\
        &= o.
    \end{aligned}
    \end{equation}
    The second line is by the definition of $q_a$.
\hfill$\blacksquare$

\medskip
\paragraph{Self-consistent PEC.} 
Formally, consider the following Self-consistent PEC (SC-PEC) protocol. One first learns a set of noise parameters $\bm\Lambda_{\bm\eta}$ that are gauge-equivalent to the true values $\bm\Lambda$, meaning that the two noise models cannot be distinguished by any experiments constructed from the noisy gate set. Assuming the learning is exact for now. Use the superscript $g$ to denote the learned noise channels. Construct our estimator using the following steps:
\begin{enumerate}
    \item Randomly sample $a_0\sim\{q_{a_0}^{\bm{\eta},S}\}$, $a_j\sim\{q^{\bm{\eta},\mc G_j}_{a_j}\}_{j=1}^T$,  $a_{T+1}\sim\{q_{a_{T+1}}^{\bm{\eta},M}\}$. 
    \item Implement and measure the following expectation value
    \begin{equation}
        E_{\bm a} = \lbra{ \widetilde O }\mc P_{a_{T+1}}\widetilde{\mc {G}}_T\mc P_{a_{T}}\mc U_T\cdots\widetilde{\mc G}_1\mc P_{a_{1}}\mc U_1\mc P_{a_{0}} \lket{\widetilde{\rho}_0},
    \end{equation}
    which is formally the same as Eq.\eqref{eq:random_sm_eq1}.
    \item Define the SC-PEC estimator as
    \begin{equation}
        \hat o^{\mr{(SC-PEC)}} = E_{\bm a}\cdot\prod_{j=0}^{T+1}\gamma^{\bm{\eta}}_j\mr{sgn}^{\bm{\eta}}_{a_j}.
    \end{equation}
    where $\gamma_j^{\bm{\eta}}$ and $\mr{sgn}^{\bm{\eta}}_{a_j}$ are with respect to the $j$th learned Pauli noise channel (instead of the true noise channel).
\end{enumerate}

The following proposition shows the correctness of SC-PEC.
\begin{proposition}[Theorem~1 in main text]
    Given that one exactly knows a $\bm\Lambda_{\bm\eta}$ that is gauge-equivalent to $\bm\Lambda$, the SC-PEC estimator with respect to $\bm\Lambda_{\bm\eta}$ is an unbiased estimator for $o$. 
\end{proposition}

\noindent\textbf{Proof.}
    First note that $E_{\bm a}$ can be expanded as
    \begin{equation}
    \begin{aligned}
        E_{\bm a} &= \lbra{ O }\Lambda_{M}\mc P_{a_{T+1}}{\mc {G}}_T\Lambda_{\mc G_T}\mc P_{a_{T}}\mc U_T\cdots{\mc G}_1\Lambda_{\mc G_1}\mc P_{a_{1}}\mc U_1\mc P_{a_{0}} \Lambda_S\lket{{\rho}_0}.
    \end{aligned}
    \end{equation}
    Since $\bm\Lambda$ and $\bm\Lambda_{\bm\eta}$ are gauge-equivalent, replacing the former with the latter by definition does not change any expectation values from any experiments. We thus have,
    \begin{equation}
        E_{\bm a} =\lbra{ O }\Lambda_{M,\bm{\eta}}\mc P_{a_{T+1}}{\mc {G}}_T\Lambda_{\mc G_T,\bm{\eta}}\mc P_{a_{T}}\mc U_T\cdots{\mc G}_1\Lambda_{\mc G_1,\bm{\eta}}\mc P_{a_{1}}\mc U_1\mc P_{a_{0}} \Lambda_{S,\bm{\eta}}\lket{{\rho}_0}.
    \end{equation}
    Then, following exactly the same argument as the proof of Proposition~\ref{prop:std-pec}, one can obtain that
    \begin{equation}
        \E {\hat o^{\mr{(SC-PEC)}}} = \lbra O \mc G_T\mc U_T\cdots\mc G_1\mc U_1 \lket{\rho_0}= o.
    \end{equation}
This completes the proof. \hfill$\blacksquare$

\subsection{Generalizations to Zero-Noise Extrapolation (ZNE)}\label{sm:ZNE}

The self-consistent formalism can be integrated with other quantum error mitigation methods besides PEC. Here, we discuss one example of self-consistent zero-noise extrapolation (ZNE). ZNE is a popular error mitigation method widely applied in experiments (e.g. \cite{kim2023evidence}). At a high level, it works by intentionally amplifying all noise channels by a few different factors (i.e., $\Lambda\mapsto\Lambda^{1+\alpha}$), estimating the observable expectation values at different noise levels, and extrapolating to the zero-noise point using a certain fitting model. For the last step, one simple choice of models is single exponential decay, but it does not always yield provably unbiased estimators~\cite{kim2023evidence}. More complicated extrapolation models can be used at the price of potential numerical instability and difficulty in implementation.

\begin{figure}[!htp]
    \centering
    \includegraphics[width=0.8\linewidth]{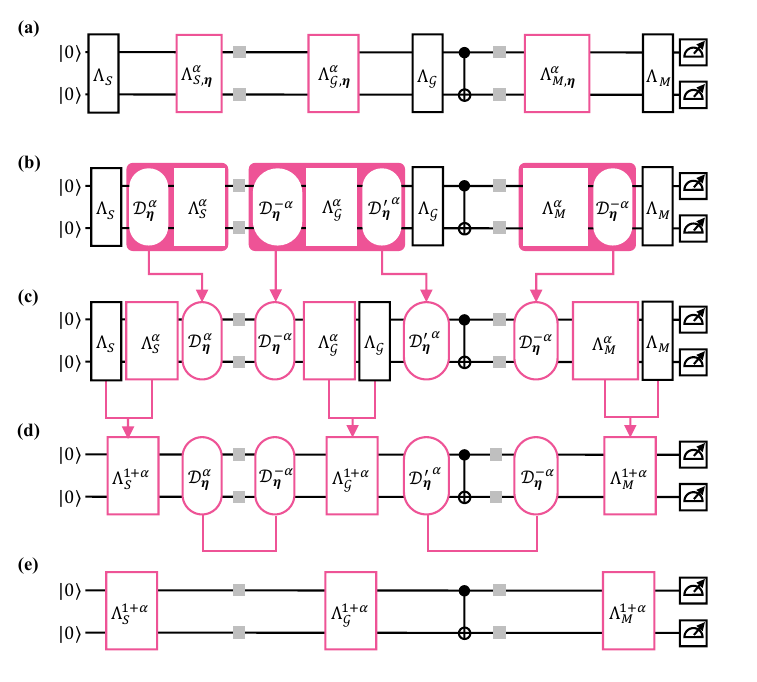}
    \caption{\textbf{Graphical proof of self-consistent ZNE.} The only difference from Figure~\ref{fig:4} is that we implement $\Lambda^{\alpha}$ for general $\alpha$ instead of $\Lambda_{\bm\eta}^{-1}$ as in PEC. (a) Implementation of $\Lambda_{\bm\eta}^\alpha$ via random Pauli gates. (b) Rewrite the noise channels using the ground truth and the gauges. (c) Commute the noise channels through. Note that both the noises and the gauges are Pauli diagonal maps, and thus commute with one another. (d) Merge the noise channels. (e) Cancel the gauge maps. Recall that $D_{\bm\eta}'$ is defined according to $\mc G\circ\mc D_{\bm\eta}' =\mc D_{\bm\eta}\circ\mc G$.}
    \label{fig:ZNE}
\end{figure}

The implementation of ZNE in~\cite{kim2023evidence} is based on probabilistic error amplification (PEA). That is, using random Pauli gates to implement a Pauli noise channel $\Lambda^\alpha$
\cite{seif_single_2026}, so as to realize arbitrary noise amplification. For this purpose, one needs to first characterize all the noise channels. We have seen that such characterization can only be done up to gauge degrees of freedom. Fortunately, once we learn a gauge-equivalent noise model $\bm\Lambda_{\bm\eta}$, PEC can be conducted self-consistently. To see this, consider again the following noisy expectation values
\begin{equation}
    o^{(\mr{noisy})} = \lbra{ \widetilde O }\widetilde{\mc {G}}_T\mc U_T\cdots\widetilde{\mc G}_1\mc U_1 \lket{\widetilde{\rho}_0}
    =\lbra O \Lambda_M\mc G_T\Lambda_{\mc G_T}\mc U_T\cdots\mc G_1\Lambda_{\mc G_1}\mc U_1 \Lambda_S\lket{\rho_0}.
\end{equation}
Suppose we exactly know a gauge-equivalent noise model $\bm\Lambda_\eta$ to the ground truth $\bm\Lambda$. We can simply implement PEA with respect to the learned model to get
\begin{equation}\label{eq:ZNE_explain}
\begin{aligned}
    o^{(\mr{PEA},\,\alpha)} &=\lbra O (\Lambda_{M,\bm\eta}^\alpha\Lambda_M)\mc G_T(\Lambda_{\mc G_T}\Lambda_{\mc G_T,\bm\eta}^\alpha)\mc U_T\cdots\mc G_1(\Lambda_{\mc G_1}\Lambda_{\mc G_1,\bm\eta}^\alpha)\mc U_1 (\Lambda_{S,\bm\eta}^\alpha\Lambda_S)\lket{\rho_0}
    \\&= \lbra O \Lambda_{M,\bm\eta}^{1+\alpha}\mc G_T\Lambda_{\mc G_T,\bm\eta}^{1+\alpha}\mc U_T\cdots\mc G_1\Lambda_{\mc G_1,\bm\eta}^{1+\alpha}\mc U_1 \Lambda_{S,\bm\eta}^{1+\alpha}\lket{\rho_0}
    \\&= \lbra O \Lambda_{M}^{1+\alpha}\mc G_T\Lambda_{\mc G_T}^{1+\alpha}\mc U_T\cdots\mc G_1\Lambda_{\mc G_1}^{1+\alpha}\mc U_1 \Lambda_{S}^{1+\alpha}\lket{\rho_0}.
\end{aligned}
\end{equation}
In the first line, we implement $\Lambda_{\bm\eta}^\alpha$ using random Pauli gates. The second line uses the definition of gauge-equivalent noise model to replace all $\Lambda$ with $\Lambda_{\bm\eta}$ without changing the expectation values. The last line says this value will be gauge-independent. To see this, recall from the main text that that all gauges in a Pauli noise model can be parameterized as~\cite{chen2026efficient}:
\begin{align}
    &\Lambda^S\mapsto\Lambda_{\bm\eta}^S=\mc D_{\bm\eta}\circ\Lambda^S,\\ &\Lambda^M\mapsto\Lambda^M_{\bm\eta}=\Lambda^M\circ\mc D_{\bm\eta}^{-1},\\ & \Lambda^{\mc G}\mapsto\Lambda^{\mc G}_{\bm\eta}= \mc D_{\bm\eta}'\circ\Lambda^{\mc G}\circ\mc D_{\bm\eta}^{-1}.
\end{align}
Further note that all the above are Pauli diagonal maps that commute with each other. We thus have, for example, $(D_{\bm\eta}'\circ\Lambda^{\mc G}\circ\mc D_{\bm\eta}^{-1})^{1+\alpha} = {D_{\bm\eta}'}^{1+\alpha}\circ{\Lambda^{\mc G}}^{1+\alpha}\circ\mc D_{\bm\eta}^{-(1+\alpha)}$. Putting these back into Eq.~\eqref{eq:ZNE_explain} and canceling the gauge maps yield the last line. This can also be seen via a graphical proof as in Fig.~\ref{fig:ZNE}, similar to our SC-PEC proof in the main text.
In conclusion, we show that PEA (and thus ZNE) can be conducted self-consistently within our framework. In particular, the noise-amplified observable expectation values $o^{(\mr{PEA},\,\alpha)}$ are gauge-independent. Still, different choice of gauge leads to different PEA implementations, which may in turn lead to different variance and mitigation cost. Conducting gauge optimization for ZNE is an interesting open problem for future exploration.

\section{How to learn self-consistently}
\subsection{Details of the quasi-local model}\label{supp:quasi-model}

In this section, we provide additional details about the quasi-local Pauli noise model.

Let us first introduce the notion of factor sets. Let $\Omega$ be a subset of $2^{[n]}$, i.e., the power set of $[n]=\{1,2,\cdots,n\}$. We call $\Omega$ a factor set if for every $\kappa\in\Omega$, every subset of $\kappa$ also belongs to $\Omega$. An exemplary factor set on $n=3$ qubits is given by $\Omega=\{\{1\},\{2\},\{3\},\{1,2\},\{2,3\}\}$. For any nontrivial $P_a\in\Pn$, we say $a\sim\Omega$ if the Pauli support of $a$ belongs to $\Omega$.
The set of all nontrivial Pauli operators given by $\Omega$ is denoted by 
\begin{equation}\label{eq:def_K}
    \mc K=\{a\sim\Omega:\forall a\in\Pn,\,a\neq0\}.
\end{equation}
In the above example, $XYI\in\mc K$ while $ZIX\not\in\mc K$. 

Recall that a Pauli channel is $\Omega$-local if it can be expressed as
\begin{equation}
    \Lambda(\cdot)=\mathop{\bigcirc}_{a\in\mc K}(\omega_a P_a(\cdot) P_a+(1-\omega_k)(\cdot)),
\end{equation}
with $\omega_a<1/2$ and we define $\tau_a = -\log(1-2\omega_a)$. 
For any $\mc K$  defined via Eq.~\eqref{eq:def_K} with a valid factor set $\Omega$, the following relations are known (Eq.~\eqref{eq:model_to_lind} in the main text),

\begin{equation}
    {x_a=\sum_{b\in\mc K}\expval{a,b}\tau_b.}
\end{equation}

The inverse transformation will become clear in a moment.

 Eq.~\eqref{eq:def_K}.  \cite[Appendix E]{chen2026efficient}.

\medskip
For the convenience of discussion, let us introduce another equivalent parameterization of $\Omega$-local Pauli channels.
For any two Pauli $a,b\in\Pn$, we write $a\triangleleft b$ if $\mr{supp}(a)\subseteq\mr{supp}(b)$ and that $a,b$ commutes at every qubit. For example, $XIYI\triangleleft XZYI$, while $XXZI \ntriangleleft YYZZ$.

Define $\bm r =\{r_a\}_{a\in\mc K}$ according to
\begin{equation}\label{eqn:mobius}
\left\{\begin{aligned}
    x_a &= \sum_{b\in\mc K:~b\triangleleft a}r_b,\quad&&\forall a\in\Pn, a\neq0,\\
    r_b &= \sum_{a\in\mc K:~a\triangleleft b}(-1)^{|b|-|a|}x_a,\quad&&\forall b\in\mc K.
\end{aligned}\right.
\end{equation}
This is known as the M\"obius transform~\cite{wagner2023learning}. 
We again note that $\mc K$ must be defined via a valid factor set for the above to hold.
$\bm r$ is referred to as the reduced parameter in Ref.~\cite{chen2026efficient}. 
The main advantages of using $\bm r$ is that every log eigenvalues $x_a=-\log\lambda_a$ can be very intuitively expressed in terms of $\bm r$ using the above equations.
Thus, we will use $\bm r$ when discussing experimental design.

Finally, it can be verified that the following transformation from $\bm r$ to $\bm\tau$ holds. Note again that $\mc{K}$ needs to be defined via a valid factor set:
\begin{equation}\label{eq:r2tau_theory}
    \tau_b = \sum_{\substack{c\in\mc K:\\ \mr{supp}(b)\subseteq\mr{supp}(c)}}\frac{-2}{4^{|c|}}(-1)^{\expval{b,c}}r_c,
\end{equation}
where $\mr{supp}(b)$ denote the Pauli support of $b$, i.e., the set of indexes $i$ where $b_i\neq I$. Combined with Eq.~\eqref{eqn:mobius}, one can obtain the inverse transformation from $\bm x$ to $\bm\tau$. All of these relations can be verified by direct substitution. They can also be found in Ref.~\cite[Appendix E]{chen2026efficient}.

\subsection{Constructing the design matrix}\label{supp:constructing_design_matrix}
Recall from Sec.~\ref{sec:model_learning} that the design matrix $F$ encodes all experimental measurements. The matrix represents a linear map between the vector of observables $\bm b$ and $\bm x$, a vector of all (log) fidelities which varies in size depending on the gate set, the number of qubits, \textbf{and} the underlying locality of the noise. For convenience, we reproduce the key expression here:
\begin{equation}\label{eq:design_matrix_sm}
    \bm b = F\bm x,
\end{equation}
where $b_j=-\log{\expval{\tilde o_j}}$ is the (log) expectation value of the $j$-th experiment.

\begin{figure}[h!]
	\centering
	\includegraphics[width=0.74\textwidth]{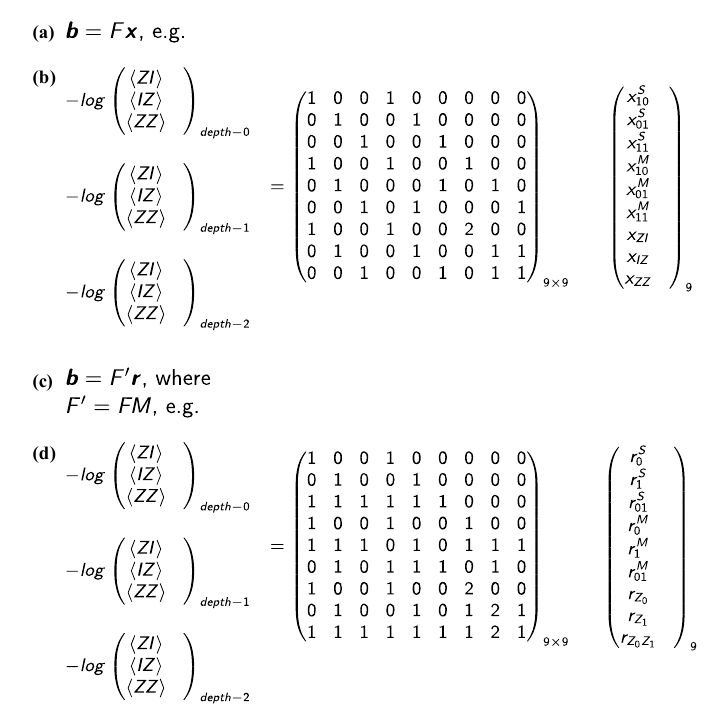} 
	\caption{
\textbf{(a)} The general expression used to infer the (log) fidelity parameters of the noise model, $\bm x$, based on experimental observables $\bm b$. The experiments are linear combinations of state preparation, gates, and measurements fidelities, and are encoded into the design matrix $F$. 
\textbf{(b)} For example, we show the design of 2Q experiments limited to observables containing only Pauli Z terms. In this restricted noise model, there are only 9 noise terms ($|\bm x|=9$), which can be inferred from depth-0, depth-1, and depth-2 (or more generally, a series of depth-even) experiments. The matrix rank of $F$ in this case is 6, which is $|\bm x|=9-\text{dim(gauge)}=9-3$, as expected.
\textbf{(c)} Although we present the design matrix $F$ in the $\bm x$ fidelity basis within the theory section of the main text and also use it for the 2Q and GHZ experiments, we also draw attention to a definition in the $\bm r$ or noise parameter basis yielding $F’$, where $\bm x=M\bm r$, and $M$ is the M\"obius transformation defined in Eq.~\eqref{eqn:mobius}. Defining $F’$ in the $\bm r$ basis allows us to keep the number of noise parameters polynomial, as opposed to exponential, in number when we impose a quasi-local model for larger system sizes.
\textbf{(d)} For example, we also present the $F’$ in the $\bm r$ basis for the 2Q experiments here.}
	\label{sfig:design_2Q}
\end{figure}

For the 2Q experiments shown in Fig.~\ref{fig:5} of the main text, it was not necessary to learn \textit{all} noise parameters if the observable being mitigated is restricted to a certain type -- in that case the $Z$-only observables. This \textit{restricted} set of noise parameters were sufficient and complete as seen in Fig.~\ref{sfig:design_2Q}b, where $Z$-only observables on $n=2$ qubits for depth-0, depth-1, and depth-2 (or more depth-even experiments) can saturate all learnable degrees of freedom subject to the remaining $2^n-1$ gauge degrees of freedom. 

We use this opportunity to describe the same analysis in a more practical noise parameterization written as $\bm r$, where the transformation $\bm x = M \bm r$ was defined earlier as the M\"obius transformation seen in Eq.~\eqref{eqn:mobius}. We show how the design matrix in the $\bm r$ basis, $F'$, is only slightly modified (Fig.~\ref{sfig:design_2Q}d) without any change, in this special case, to the number of parameters in the noise model ($|\bm x| = |\bm r | $). In this more convenient basis, the design matrix $F'$ can be seen to be complete as long as the matrix rank of $F'$ is equal to $|\bm r| - (2^n-1)$ for a general noise model, or $|\bm r| - n$ for two-local noise model that only admits single-qubit gauge transformation (e.g. Fig.~\ref{fig:3})~\cite{chen2026efficient}.

\begin{figure}
	\centering
	\includegraphics[width=0.74\textwidth]{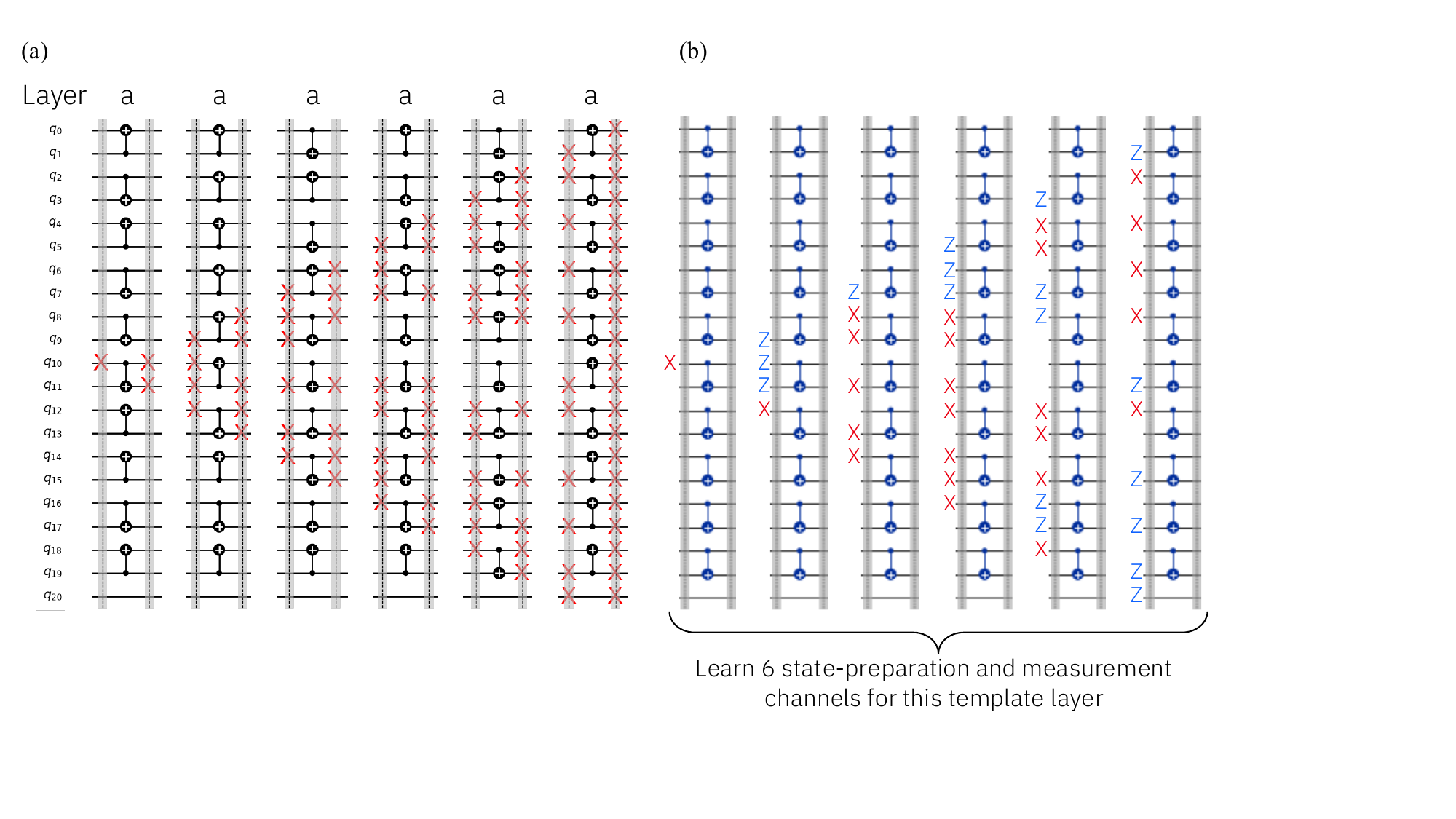} 
	\caption{For one of the two template layers $a$ and $b$ used for preparing the $n=21$ GHZ state, we looked at those Pauli eigenvalues in (\textbf{a}) which contribute to the $\langle X^n \rangle$ observable, relying on the fact that this is a Clifford circuit and thus back-propagation was possible. (\textbf{b}) Since the template layer $a$ may not have the CNOT in the same directions as the seen in the target circuit, the input and output bases must be properly accounted for after pre- and post-pending single-qubit Hadamard gates around the gate, thus the conversion from $X$-only input to some $Z$-type Paulis. For this template, six state preparation and measurement channels were needed. We performed the same task for template layer $b$, and grouped these experiments, commuting-wise, after appending the preparation and measurement bases into a single string of length $2n$. This allowed us to reduce the number of preparation and learning experiments down to 7, depth-1, and 7 depth-even experiments to invert the design matrix $F$ of dimensions $(56\times46)$.}\label{sfig:ghz_learning}
\end{figure}

For larger system sizes $n$ even with a \textit{restricted} noise model, the number of learning experiments not only depends on the Hamming weight of the target observable, but also grows rapidly with the system size itself. In the case of the largest GHZ state we prepared on $n=21$ qubits, the number of gauge parameters is, in theory, as large as $2^{21}-1$. However, because we took advantage of the fact that the target circuit is a Clifford circuit, whose final observable could be classically back-propagated, we focused our experiments exclusively on learning those noise fidelities $\bm x$ which contributed to corrupting the target observable $\langle X^n\rangle$ (See Fig.~\ref{sfig:ghz_learning} for an illustration of this procedure on one of the two template layers). To be exact, the $n=21$ GHZ state required: a single, depth-0 experiment for SPAM, 7 depth-1 experiments, $7$ depth-even experiments for each depths-even circuits of 2, 4, and 8. In total, 29 learning experiments informed the 56 observables needed to unambiguously infer the 46 fidelity terms in $\bm x$. The inferred noise model was used to predict $\langle X^n\rangle$, and compared against the experimentally measured value for the target circuit (Fig.~\ref{fig:6}). Although we performed this analysis in the $\bm x$ basis (as opposed to the $\bm r$ basis), we verified that the design matrix $F$ was complete by observing that rank($F$), 34, and the number of SPAM bases, 12, add up to the total number of unknown fidelities $|\bm x |=46$ (See Table~\ref{tab:experiment_settings}). This same procedure was used, with overlapping experiments where possible, for all the GHZ system sizes from $n=3$ to $n=21$.

\small
\begin{table}
    \fontfamily{pcr}\selectfont
    \centering
    \scalebox{0.7}
    {
    \begin{tabular}{|c|c|c|}
    \hline
    Count & Depth/Type & Experiment \\
    \hline
\hline 1 & Depth-even & $2*x^1_{IIIIIIIXXXIZZIIIIIIII} + 2*x^1_{IIIIIIXXIXIZZZIIIIIII} + x^m_{IIIIIIIZZZIZZIIIIIIII} + x^s_{IIIIIIIZZZIZZIIIIIIII}$ \\
\hline 2 & Depth-1 & $x^1_{IXXXIZIIZIIIZZXIZZZII} + x^m_{ZZIZIZIIZZIIZIZIZIZZI} + x^s_{IZZZIZIIZIIIZZZIZZZII}$ \\
\hline 3 & Depth-even & $x^1_{IIIIIIIXXXIZZIIIIIIII} + x^1_{IIIIIIXXIXIZZZIIIIIII} + x^m_{IIIIIIIZZZIZZIIIIIIII} + x^s_{IIIIIIIZZZIZZIIIIIIII}$ \\
\hline 4 & Depth-1 & $x^1_{IIIXZIXXZZIIIXIZZIIII} + x^m_{IIZZZZIZZIIIZZIZZZIII} + x^s_{IIIZZIZZZZIIIZIZZIIII}$ \\
\hline 5 & Depth-even & $2*x^2_{ZZIZIZIIXZIIXIXIXIXZI} + 2*x^2_{ZZZZZZZXXZZXXXXXXXXZZ} + x^m_{ZZIZIZIIZZIIZIZIZIZZI} + x^s_{ZZIZIZIIZZIIZIZIZIZZI}$ \\
\hline 6 & Depth-even & $2*x^2_{IIIIIIXXIXIXXZIIIIIII} + 2*x^2_{IIIIIXXXIXIIXZZIIIIII} + x^m_{IIIIIIZZIZIZZZIIIIIII} + x^s_{IIIIIIZZIZIZZZIIIIIII}$ \\
\hline 7 & Depth-even & $2*x^2_{IIIIXXXIXXIIIZZZIIIII} + 2*x^2_{IIIXXIXXXXIIIZIZZIIII} + x^m_{IIIIZZZIZZIIIZZZIIIII} + x^s_{IIIIZZZIZZIIIZZZIIIII}$ \\
\hline 8 & Depth-even & $2*x^2_{IIIIIIIIIIXIIIIIIIIII} + 2*x^2_{IIIIIIIIIXXIIIIIIIIII} + x^m_{IIIIIIIIIIZIIIIIIIIII} + x^s_{IIIIIIIIIIZIIIIIIIIII}$ \\
\hline 9 & Depth-even & $2*x^1_{IIIIIXZZIXIIXXZIIIIII} + 2*x^1_{IIIIXXZIXXIIIXZZIIIII} + x^m_{IIIIIZZZIZIIZZZIIIIII} + x^s_{IIIIIZZZIZIIZZZIIIIII}$ \\
\hline 10 & Depth-even & $2*x^1_{IIIIIIIIIXZIIIIIIIIII} + 2*x^1_{IIIIIIIIXXZZIIIIIIIII} + x^m_{IIIIIIIIIZZIIIIIIIIII} + x^s_{IIIIIIIIIZZIIIIIIIIII}$ \\
\hline 11 & Depth-even & $4*x^1_{IIIXZIXXZZIIIXIZZIIII} + 4*x^1_{IIXXZZIXZIIIXXIZZZIII} + x^m_{IIIZZIZZZZIIIZIZZIIII} + x^s_{IIIZZIZZZZIIIZIZZIIII}$ \\
\hline 12 & Depth-even & $2*x^1_{IIIXZIXXZZIIIXIZZIIII} + 2*x^1_{IIXXZZIXZIIIXXIZZZIII} + x^m_{IIIZZIZZZZIIIZIZZIIII} + x^s_{IIIZZIZZZZIIIZIZZIIII}$ \\
\hline 13 & Depth-even & $4*x^2_{IIIIIIIIXZZZIIIIIIIII} + 4*x^2_{IIIIIIIXXZIZZIIIIIIII} + x^m_{IIIIIIIIZZZZIIIIIIIII} + x^s_{IIIIIIIIZZZZIIIIIIIII}$ \\
\hline 14 & Depth-even & $4*x^2_{IIXZZXIXXIIIZZIXXZIII} + 4*x^2_{IXXZIXIIXIIIZZZIXZZII} + x^m_{IIZZZZIZZIIIZZIZZZIII} + x^s_{IIZZZZIZZIIIZZIZZZIII}$ \\
\hline 15 & Depth-even & $x^2_{IIIIIIXXIXIXXZIIIIIII} + x^2_{IIIIIXXXIXIIXZZIIIIII} + x^m_{IIIIIIZZIZIZZZIIIIIII} + x^s_{IIIIIIZZIZIZZZIIIIIII}$ \\
\hline 16 & Depth-even & $x^2_{IIIIXXXIXXIIIZZZIIIII} + x^2_{IIIXXIXXXXIIIZIZZIIII} + x^m_{IIIIZZZIZZIIIZZZIIIII} + x^s_{IIIIZZZIZZIIIZZZIIIII}$ \\
\hline 17 & Depth-even & $x^2_{IIIIIIIIIIXIIIIIIIIII} + x^2_{IIIIIIIIIXXIIIIIIIIII} + x^m_{IIIIIIIIIIZIIIIIIIIII} + x^s_{IIIIIIIIIIZIIIIIIIIII}$ \\
\hline 18 & Depth-even & $4*x^1_{IIIIIIIXXXIZZIIIIIIII} + 4*x^1_{IIIIIIXXIXIZZZIIIIIII} + x^m_{IIIIIIIZZZIZZIIIIIIII} + x^s_{IIIIIIIZZZIZZIIIIIIII}$ \\
\hline 19 & Depth-1 & $x^2_{ZZIZIZIIXZIIXIXIXIXZI} + x^m_{ZZZZZZZZZZZZZZZZZZZZZ} + x^s_{ZZIZIZIIZZIIZIZIZIZZI}$ \\
\hline 20 & Depth-1 & $x^2_{IIIIIIIIXZZZIIIIIIIII} + x^m_{IIIIIIIZZZIZZIIIIIIII} + x^s_{IIIIIIIIZZZZIIIIIIIII}$ \\
\hline 21 & Depth-1 & $x^2_{IIXZZXIXXIIIZZIXXZIII} + x^m_{IZZZIZIIZIIIZZZIZZZII} + x^s_{IIZZZZIZZIIIZZIZZZIII}$ \\
\hline 22 & Depth-1 & $x^2_{IIIIIIXXIXIXXZIIIIIII} + x^m_{IIIIIZZZIZIIZZZIIIIII} + x^s_{IIIIIIZZIZIZZZIIIIIII}$ \\
\hline 23 & Depth-1 & $x^2_{IIIIXXXIXXIIIZZZIIIII} + x^m_{IIIZZIZZZZIIIZIZZIIII} + x^s_{IIIIZZZIZZIIIZZZIIIII}$ \\
\hline 24 & Depth-1 & $x^2_{IIIIIIIIIIXIIIIIIIIII} + x^m_{IIIIIIIIIZZIIIIIIIIII} + x^s_{IIIIIIIIIIZIIIIIIIIII}$ \\
\hline 25 & Depth-even & $x^2_{ZZIZIZIIXZIIXIXIXIXZI} + x^2_{ZZZZZZZXXZZXXXXXXXXZZ} + x^m_{ZZIZIZIIZZIIZIZIZIZZI} + x^s_{ZZIZIZIIZZIIZIZIZIZZI}$ \\
\hline 26 & Depth-even & $4*x^2_{ZZIZIZIIXZIIXIXIXIXZI} + 4*x^2_{ZZZZZZZXXZZXXXXXXXXZZ} + x^m_{ZZIZIZIIZZIIZIZIZIZZI} + x^s_{ZZIZIZIIZZIIZIZIZIZZI}$ \\
\hline 27 & Depth-even & $x^2_{IIIIIIIIXZZZIIIIIIIII} + x^2_{IIIIIIIXXZIZZIIIIIIII} + x^m_{IIIIIIIIZZZZIIIIIIIII} + x^s_{IIIIIIIIZZZZIIIIIIIII}$ \\
\hline 28 & Depth-even & $x^2_{IIXZZXIXXIIIZZIXXZIII} + x^2_{IXXZIXIIXIIIZZZIXZZII} + x^m_{IIZZZZIZZIIIZZIZZZIII} + x^s_{IIZZZZIZZIIIZZIZZZIII}$ \\
\hline 29 & Depth-even & $4*x^2_{IIIIIIXXIXIXXZIIIIIII} + 4*x^2_{IIIIIXXXIXIIXZZIIIIII} + x^m_{IIIIIIZZIZIZZZIIIIIII} + x^s_{IIIIIIZZIZIZZZIIIIIII}$ \\
\hline 30 & Depth-even & $4*x^2_{IIIIXXXIXXIIIZZZIIIII} + 4*x^2_{IIIXXIXXXXIIIZIZZIIII} + x^m_{IIIIZZZIZZIIIZZZIIIII} + x^s_{IIIIZZZIZZIIIZZZIIIII}$ \\
\hline 31 & Depth-even & $4*x^2_{IIIIIIIIIIXIIIIIIIIII} + 4*x^2_{IIIIIIIIIXXIIIIIIIIII} + x^m_{IIIIIIIIIIZIIIIIIIIII} + x^s_{IIIIIIIIIIZIIIIIIIIII}$ \\
\hline 32 & Depth-even & $2*x^2_{IIIIIIIIXZZZIIIIIIIII} + 2*x^2_{IIIIIIIXXZIZZIIIIIIII} + x^m_{IIIIIIIIZZZZIIIIIIIII} + x^s_{IIIIIIIIZZZZIIIIIIIII}$ \\
\hline 33 & Depth-even & $2*x^2_{IIXZZXIXXIIIZZIXXZIII} + 2*x^2_{IXXZIXIIXIIIZZZIXZZII} + x^m_{IIZZZZIZZIIIZZIZZZIII} + x^s_{IIZZZZIZZIIIZZIZZZIII}$ \\
\hline 34 & Depth-even & $x^1_{IIIXZIXXZZIIIXIZZIIII} + x^1_{IIXXZZIXZIIIXXIZZZIII} + x^m_{IIIZZIZZZZIIIZIZZIIII} + x^s_{IIIZZIZZZZIIIZIZZIIII}$ \\
\hline 35 & Depth-1 & $x^1_{IIIIIXZZIXIIXXZIIIIII} + x^m_{IIIIZZZIZZIIIZZZIIIII} + x^s_{IIIIIZZZIZIIZZZIIIIII}$ \\
\hline 36 & Depth-1 & $x^1_{IIIIIIIIIXZIIIIIIIIII} + x^m_{IIIIIIIIZZZZIIIIIIIII} + x^s_{IIIIIIIIIZZIIIIIIIIII}$ \\
\hline 37 & Depth-even & $x^1_{IXXXIZIIZIIIZZXIZZZII} + x^1_{XXIXIZIIZZIIZIXIZIZZI} + x^m_{IZZZIZIIZIIIZZZIZZZII} + x^s_{IZZZIZIIZIIIZZZIZZZII}$ \\
\hline 38 & Depth-even & $2*x^1_{IXXXIZIIZIIIZZXIZZZII} + 2*x^1_{XXIXIZIIZZIIZIXIZIZZI} + x^m_{IZZZIZIIZIIIZZZIZZZII} + x^s_{IZZZIZIIZIIIZZZIZZZII}$ \\
\hline 39 & Depth-even & $4*x^1_{IIIIIXZZIXIIXXZIIIIII} + 4*x^1_{IIIIXXZIXXIIIXZZIIIII} + x^m_{IIIIIZZZIZIIZZZIIIIII} + x^s_{IIIIIZZZIZIIZZZIIIIII}$ \\
\hline 40 & Depth-even & $4*x^1_{IIIIIIIIIXZIIIIIIIIII} + 4*x^1_{IIIIIIIIXXZZIIIIIIIII} + x^m_{IIIIIIIIIZZIIIIIIIIII} + x^s_{IIIIIIIIIZZIIIIIIIIII}$ \\
\hline 41 & Depth-even & $4*x^1_{IXXXIZIIZIIIZZXIZZZII} + 4*x^1_{XXIXIZIIZZIIZIXIZIZZI} + x^m_{IZZZIZIIZIIIZZZIZZZII} + x^s_{IZZZIZIIZIIIZZZIZZZII}$ \\
\hline 42 & Depth-even & $x^1_{IIIIIXZZIXIIXXZIIIIII} + x^1_{IIIIXXZIXXIIIXZZIIIII} + x^m_{IIIIIZZZIZIIZZZIIIIII} + x^s_{IIIIIZZZIZIIZZZIIIIII}$ \\
\hline 43 & Depth-even & $x^1_{IIIIIIIIIXZIIIIIIIIII} + x^1_{IIIIIIIIXXZZIIIIIIIII} + x^m_{IIIIIIIIIZZIIIIIIIIII} + x^s_{IIIIIIIIIZZIIIIIIIIII}$ \\
\hline 44 & Depth-1 & $x^1_{IIIIIIIXXXIZZIIIIIIII} + x^m_{IIIIIIZZIZIZZZIIIIIII} + x^s_{IIIIIIIZZZIZZIIIIIIII}$ \\
\hline 45 & SPAM & $x^m_{IIIIIIZZIZIZZZIIIIIII} + x^s_{IIIIIIZZIZIZZZIIIIIII}$ \\
\hline 46 & SPAM & $x^m_{IIIIIIIIZZZZIIIIIIIII} + x^s_{IIIIIIIIZZZZIIIIIIIII}$ \\
\hline 47 & SPAM & $x^m_{IIIIIIIIIIZIIIIIIIIII} + x^s_{IIIIIIIIIIZIIIIIIIIII}$ \\
\hline 48 & SPAM & $x^m_{IZZZIZIIZIIIZZZIZZZII} + x^s_{IZZZIZIIZIIIZZZIZZZII}$ \\
\hline 49 & SPAM & $x^m_{IIIIIZZZIZIIZZZIIIIII} + x^s_{IIIIIZZZIZIIZZZIIIIII}$ \\
\hline 50 & SPAM & $x^m_{ZZZZZZZZZZZZZZZZZZZZZ} + x^s_{ZZZZZZZZZZZZZZZZZZZZZ}$ \\
\hline 51 & SPAM & $x^m_{ZZIZIZIIZZIIZIZIZIZZI} + x^s_{ZZIZIZIIZZIIZIZIZIZZI}$ \\
\hline 52 & SPAM & $x^m_{IIIIIIIZZZIZZIIIIIIII} + x^s_{IIIIIIIZZZIZZIIIIIIII}$ \\
\hline 53 & SPAM & $x^m_{IIIZZIZZZZIIIZIZZIIII} + x^s_{IIIZZIZZZZIIIZIZZIIII}$ \\
\hline 54 & SPAM & $x^m_{IIZZZZIZZIIIZZIZZZIII} + x^s_{IIZZZZIZZIIIZZIZZZIII}$ \\
\hline 55 & SPAM & $x^m_{IIIIIIIIIZZIIIIIIIIII} + x^s_{IIIIIIIIIZZIIIIIIIIII}$ \\
\hline 56 & SPAM & $x^m_{IIIIZZZIZZIIIZZZIIIII} + x^s_{IIIIZZZIZZIIIZZZIIIII}$ \\
\hline
        \end{tabular}
        }
\caption{
To mitigate the $n=21$ GHZ experiment, we required a total of 56 learning experiments (rows of the design matrix) containing a total of 46 unique fidelity parameters (columns of the design matrix). The fidelity parameters $\lambda$ with superscripts `$s$' denote state-preparation, `$m$' for measurement, `$1$' for layer-1, and `$2$' for layer-2. The Pauli eigenvalues are the subscripts ordered from qubits 1 to 21, from left to right.
}
    \label{tab:ghz}
\end{table}

To move beyond \textit{restricted} noise models, we needed to impose locality in the noise so that the number of noise parameters did not continue to grow exponentially in system size. For this task, we utilized the design principle outlined in~\cite{chen2026efficient}, and also briefly discussed throughout the sections above. Unlike the previous two examples, knowledge of the target observable was not used to inform the creation of the design matrix $F'$ (in this case $\bm r$ basis) -- instead, we conduct a \emph{complete} learning of the quasi-local noise model. To measure the $9,108$ rows of observables for estimating all $2,576$ noise parameters in $\bm r$, we needed a total of 1 circuit for SPAM, 17 circuits for each template layer at depth-1, and 9 circuits for each template layer for multiple depth-even values (e.g. 4, 12, and 24). The explicit input and output bases can be found in Table~\ref{tab:ring_bases}, and the additional details in Table~\ref{tab:experiment_settings}.

To characterize all the learnable parameters to additive precision it suffices to only perform a single, depth-0 SPAM experiment \textit{and} additional depth-1 experiments for each layer. The depth-1 experiments involve the preparation of a Pauli eigenstate, a single application of the layer, terminated by a measurement in a Pauli basis that can be different than the initial one.

However, in the low-error regime, it is desirable to learn the parameters with multiplicative precision, which means the estimates can be improved with repeated applications of the gates. Therefore, we augment these experiments with additional even-depth experiments involving the preparation of a Pauli eigenstate, an even number of applications of the layer, and measurements in the \textit{same} Pauli basis for a local two-qubit basis (9 experiments per depth per layer). Additionally, when the input and output Paulis commute qubit-wise, the corresponding experiments can be combined. With these considerations, we reduced the total circuit count for both layers from 54 to 34 for the depth-1 experiments, and from 38 to 18 for the depth-even experiments (See Table~\ref{tab:ring_bases} for the exact input- and output-bases). 

In our learning experiments, we measured depths of 4, 12, and 24 for the depth-even learning experiments. We emphasize that the number of experiments we have designed does not depend on the number of qubits, or the size of the qubit ring as long it is a multiple of four. Finally, to ensure numerical stability in inferring the model parameters from the \textit{logarithm} of the measured expectation values $\langle O \rangle$, the largest depth of these learning experiments need to be smaller than the inverse of the typical gate error rate, e.g. for gate errors of $\approx1\%$, circuit depths should be less than $\approx100$; also, learning circuits sampled with enough repetitions such that statistical fluctuations $\sigma_{\langle O\rangle}$ of the measured observables are much smaller than the measured outcomes $\langle O \rangle$.

\begin{table}[h!]
    \fontfamily{pcr}\selectfont
    \centering
    \begin{tabular}{|c|c|c|c|c|c|c|}
    \hline
    Count & \multicolumn{2}{|c|}{Layer 0} & \multicolumn{2}{|c|}{Layer a} & \multicolumn{2}{|c|}{Layer b} \\
    \hline
         & input & output &  input & output & input & output \\
        \hline
        \hline
SPAM &        ZZZZZZZZZZZZ & ZZZZZZZZZZZZ & \multicolumn{4}{|c|}{}\\
\hline
\multicolumn{3}{|c|}{} & \multicolumn{4}{|c|}{Depth-1} \\
\hline 1 & \multicolumn{2}{|c|}{} &YZYZYZYZYZYZ  &  XYXYXYXYXYXY  &  ZYZYZYZYZYZY  &  YXYXYXYXYXYX \\
 & \multicolumn{2}{|c|}{} &  &  \textcolor{gray}{XYXYXYXYXYXY}  &    &  \textcolor{gray}{YXYXYXYXYXYX} \\
\hline 2 & \multicolumn{2}{|c|}{} &YYYYYYYYYYYY  &  XZXZXZXZXZXZ  &  ZXZXZXZXZXZX  &  YYYYYYYYYYYY \\
 & \multicolumn{2}{|c|}{} &  &  \textcolor{gray}{XZXZXZXZXZXZ}  &    &  \textcolor{gray}{YYYYYYYYYYYY} \\
\hline 3 & \multicolumn{2}{|c|}{} &XZXZXZXZXZXZ  &  YYYYYYYYYYYY  &  YYYYYYYYYYYY  &  ZXZXZXZXZXZX \\
 & \multicolumn{2}{|c|}{} &  &  \textcolor{gray}{YYYYYYYYYYYY}  &    &  \textcolor{gray}{ZXZXZXZXZXZX} \\
\hline 4 & \multicolumn{2}{|c|}{} &XYXYXYXYXYXY  &  YZYZYZYZYZYZ  &  YXYXYXYXYXYX  &  ZYZYZYZYZYZY \\
 & \multicolumn{2}{|c|}{} &  &  \textcolor{gray}{YZYZYZYZYZYZ}  &    &  \textcolor{gray}{ZYZYZYZYZYZY} \\
\hline 5 & \multicolumn{2}{|c|}{} &ZYXXZYXXZYXX  &  ZYXXZYXXZYXX  &  XZYXXZYXXZYX  &  XZYXXZYXXZYX \\
 & \multicolumn{2}{|c|}{} &  &  \textcolor{gray}{IYXIIYXIIYXI}  &    &  \textcolor{gray}{IIYXIIYXIIYX} \\
 & \multicolumn{2}{|c|}{} &  &  \textcolor{gray}{ZIIXZIIXZIIX}  &    &  \textcolor{gray}{XZIIXZIIXZII} \\
\hline 6 & \multicolumn{2}{|c|}{} &XXZYXXZYXXZY  &  XXZYXXZYXXZY  &  YXXZYXXZYXXZ  &  YXXZYXXZYXXZ \\
 & \multicolumn{2}{|c|}{} &  &  \textcolor{gray}{XIIYXIIYXIIY}  &    &  \textcolor{gray}{YXIIYXIIYXII} \\
 & \multicolumn{2}{|c|}{} &  &  \textcolor{gray}{IXZIIXZIIXZI}  &    &  \textcolor{gray}{IIXZIIXZIIXZ} \\
\hline 7 & \multicolumn{2}{|c|}{} &ZYYXZYYXZYYX  &  IYYIIYYIIYYI  &  XZYYXZYYXZYY  &  IIYYIIYYIIYY \\
 & \multicolumn{2}{|c|}{} &  &  \textcolor{gray}{IYYIIYYIIYYI}  &    &  \textcolor{gray}{IIYYIIYYIIYY} \\
\hline 8 & \multicolumn{2}{|c|}{} &YXZYYXZYYXZY  &  YIIYYIIYYIIY  &  YYXZYYXZYYXZ  &  YYIIYYIIYYII \\
 & \multicolumn{2}{|c|}{} &  &  \textcolor{gray}{YIIYYIIYYIIY}  &    &  \textcolor{gray}{YYIIYYIIYYII} \\
\hline 9 & \multicolumn{2}{|c|}{} &ZZXXZZXXZZXX  &  IZXIIZXIIZXI  &  XZZXXZZXXZZX  &  IIZXIIZXIIZX \\
 & \multicolumn{2}{|c|}{} &  &  \textcolor{gray}{IZXIIZXIIZXI}  &    &  \textcolor{gray}{IIZXIIZXIIZX} \\
\hline 10 & \multicolumn{2}{|c|}{} &XXZZXXZZXXZZ  &  XIIZXIIZXIIZ  &  ZXXZZXXZZXXZ  &  ZXIIZXIIZXII \\
 & \multicolumn{2}{|c|}{} &  &  \textcolor{gray}{XIIZXIIZXIIZ}  &    &  \textcolor{gray}{ZXIIZXIIZXII} \\
\hline 11 & \multicolumn{2}{|c|}{} &ZZYXZZYXZZYX  &  IZYIIZYIIZYI  &  XZZYXZZYXZZY  &  IIZYIIZYIIZY \\
 & \multicolumn{2}{|c|}{} &  &  \textcolor{gray}{IZYIIZYIIZYI}  &    &  \textcolor{gray}{IIZYIIZYIIZY} \\
\hline 12 & \multicolumn{2}{|c|}{} &YXZZYXZZYXZZ  &  YIIZYIIZYIIZ  &  ZYXZZYXZZYXZ  &  ZYIIZYIIZYII \\
 & \multicolumn{2}{|c|}{} &  &  \textcolor{gray}{YIIZYIIZYIIZ}  &    &  \textcolor{gray}{ZYIIZYIIZYII} \\
\hline 13 & \multicolumn{2}{|c|}{} &XXXXXXXXXXXX  &  XXXXXXXXXXXX  &  XXXXXXXXXXXX  &  XXXXXXXXXXXX \\
 & \multicolumn{2}{|c|}{} &  &  \textcolor{gray}{XXXXXXXXXXXX}  &    &  \textcolor{gray}{XXXXXXXXXXXX} \\
 & \multicolumn{2}{|c|}{} &  &  \textcolor{gray}{IXXIIXXIIXXI}  &    &  \textcolor{gray}{IIXXIIXXIIXX} \\
 & \multicolumn{2}{|c|}{} &  &  \textcolor{gray}{XIIXXIIXXIIX}  &    &  \textcolor{gray}{XXIIXXIIXXII} \\
\hline 14 & \multicolumn{2}{|c|}{} &YXYXYXYXYXYX  &  YXYXYXYXYXYX  &  XYXYXYXYXYXY  &  XYXYXYXYXYXY \\
 & \multicolumn{2}{|c|}{} &  &  \textcolor{gray}{YXYXYXYXYXYX}  &    &  \textcolor{gray}{XYXYXYXYXYXY} \\
 & \multicolumn{2}{|c|}{} &  &  \textcolor{gray}{IXYIIXYIIXYI}  &    &  \textcolor{gray}{IIXYIIXYIIXY} \\
 & \multicolumn{2}{|c|}{} &  &  \textcolor{gray}{YIIXYIIXYIIX}  &    &  \textcolor{gray}{XYIIXYIIXYII} \\
\hline 15 & \multicolumn{2}{|c|}{} &ZXZXZXZXZXZX  &  ZXZXZXZXZXZX  &  XZXZXZXZXZXZ  &  XZXZXZXZXZXZ \\
 & \multicolumn{2}{|c|}{} &  &  \textcolor{gray}{ZXZXZXZXZXZX}  &    &  \textcolor{gray}{XZXZXZXZXZXZ} \\
\hline 16 & \multicolumn{2}{|c|}{} &ZYZYZYZYZYZY  &  ZYZYZYZYZYZY  &  YZYZYZYZYZYZ  &  YZYZYZYZYZYZ \\
 & \multicolumn{2}{|c|}{} &  &  \textcolor{gray}{ZYZYZYZYZYZY}  &    &  \textcolor{gray}{YZYZYZYZYZYZ} \\
 & \multicolumn{2}{|c|}{} &  &  \textcolor{gray}{IYZIIYZIIYZI}  &    &  \textcolor{gray}{IIYZIIYZIIYZ} \\
 & \multicolumn{2}{|c|}{} &  &  \textcolor{gray}{ZIIYZIIYZIIY}  &    &  \textcolor{gray}{YZIIYZIIYZII} \\
\hline 17 & \multicolumn{2}{|c|}{} &ZZZZZZZZZZZZ  &  ZZZZZZZZZZZZ  &  ZZZZZZZZZZZZ  &  ZZZZZZZZZZZZ \\
 & \multicolumn{2}{|c|}{} &  &  \textcolor{gray}{ZZZZZZZZZZZZ}  &    &  \textcolor{gray}{ZZZZZZZZZZZZ} \\
 & \multicolumn{2}{|c|}{} &  &  \textcolor{gray}{IZZIIZZIIZZI}  &    &  \textcolor{gray}{IIZZIIZZIIZZ} \\
 & \multicolumn{2}{|c|}{} &  &  \textcolor{gray}{ZIIZZIIZZIIZ}  &    &  \textcolor{gray}{ZZIIZZIIZZII} \\
        \hline
        \hline
\multicolumn{3}{|c|}{} & \multicolumn{4}{|c|}{Depth-even} \\
 \hline 1 & \multicolumn{2}{|c|}{} &XXXXXXXXXXXX  &  XXXXXXXXXXXX  &  XXXXXXXXXXXX  &  XXXXXXXXXXXX \\
\hline 2 & \multicolumn{2}{|c|}{} &XYXYXYXYXYXY  &  XYXYXYXYXYXY  &  XYXYXYXYXYXY  &  XYXYXYXYXYXY \\
\hline 3 & \multicolumn{2}{|c|}{} &XZXZXZXZXZXZ  &  XZXZXZXZXZXZ  &  XZXZXZXZXZXZ  &  XZXZXZXZXZXZ \\
\hline 4 & \multicolumn{2}{|c|}{} &YXYXYXYXYXYX  &  YXYXYXYXYXYX  &  YXYXYXYXYXYX  &  YXYXYXYXYXYX \\
\hline 5 & \multicolumn{2}{|c|}{} &YYYYYYYYYYYY  &  YYYYYYYYYYYY  &  YYYYYYYYYYYY  &  YYYYYYYYYYYY \\
\hline 6 & \multicolumn{2}{|c|}{} &YZYZYZYZYZYZ  &  YZYZYZYZYZYZ  &  YZYZYZYZYZYZ  &  YZYZYZYZYZYZ \\
\hline 7 & \multicolumn{2}{|c|}{} &ZXZXZXZXZXZX  &  ZXZXZXZXZXZX  &  ZXZXZXZXZXZX  &  ZXZXZXZXZXZX \\
\hline 8 & \multicolumn{2}{|c|}{} &ZYZYZYZYZYZY  &  ZYZYZYZYZYZY  &  ZYZYZYZYZYZY  &  ZYZYZYZYZYZY \\
\hline 9 & \multicolumn{2}{|c|}{} &ZZZZZZZZZZZZ  &  ZZZZZZZZZZZZ  &  ZZZZZZZZZZZZ  &  ZZZZZZZZZZZZ \\
         \hline
    \end{tabular}
    \caption{This table shows the input- and output-bases for the learning circuits needed for SPAM (1), for depth-1 observables (17 per layer) and depth-even observables (9 per layer per depth) necessary for the design matrix $F$. Those output bases in gray can be measured simultaneously using the output bases at the top of each cell. This assumes 12 qubits on a closed ring. For more qubits, which also must be a multiple of four and on a closed ring, the bases are repeated with a period of four. For other qubit topologies, such as lines or lattices, new designs must be undertaken.}
    \label{tab:ring_bases}
\end{table}

\section{Experimental details}

In this section we give further details on the experiments presented in the main text. We refer to these as the two-qubit experiment from Sec.~\ref{sec:2qubit}, the GHZ preparation experiment in Sec.~\ref{sec:GHZ}, and the ring circuit experiment in Sec.~\ref{sec:ring}.
We summarize the main differences between these experiments in Table~\ref{tab:experiment_settings}. 

For all executed circuits, we employed uniform Pauli twirling of the respective two-qubit gate layers to suppress coherent errors and justify the assumption of a Pauli noise channel. 
That is, we ran several instances of circuits, known as ``twirls'', that implement the same global unitary but differ in their single-qubit gate layers.
Despite this randomized circuit compilation overhead, we maintained kHZ sampling rates by making use of a recently introduced parametric circuit compilation and parameter binding pipeline facilitated by the \texttt{Sampler} primitive within the IBM Qiskit runtime service~\cite{qiskit2024}.
Moreover, we symmetrized the noise channel of the readout by also twirling measurements through random insertion of Pauli $\hat X$ or $\hat I$ gates (sampled uniformly) prior to the readout~\cite{van2022model}.
Nonetheless, the overhead of running different twirling and measurement configurations remains non-negligible, which is why we collected multiple measurements (``shots'') for each twirled circuit (See Table~\ref{tab:experiment_settings}). 

For each experiment, we learned both a ``inconsistent'' noise model and a self-consistent noise model. 
The inconsistent models derive from the learning theory originally established in Ref.~\cite{van2023probabilistic}: For each noisy layer, we implemented a given number of even-depth learning circuits for a basis of Paulis as specified in Table~\ref{tab:experiment_settings}.
In this context, SPAM errors were dealt with independently from gate noise following the technique from Ref.~\cite{van2022model} known also as \emph{twirled readout error extinction} (TREX). That is, the noisy expectation value of an observable $O$ was divided by an estimate of $\bra{0}O\ket{0}$ in a prepare-$\ket{0}$ circuit (under measurement twirling). 
The noisy estimate of $\bra{0}O\ket{0}$ was performed with the same number of twirls and shots per twirl as stated in Table~\ref{tab:experiment_settings}.
Finally, the set of learning circuits for the self-consistent noise models comprises the same even-depth learning circuits used for the inconsistent model as well as additional depth-one learning circuits for the respective Pauli basis of the model.

\begin{table}
\begin{center}
\begin{tblr}{|l|c|c|c|c|}
\hline 
Experiment  & {single CNOT \\ (see Sec.~\ref{sec:2qubit})}  &  {GHZ preparation \\ (see Sec.~\ref{sec:GHZ})} & {ring circuit \\ (see Sec.~\ref{sec:ring})} & {non-Clifford circuit \\ (see Sec.~\ref{sec:gauge-optimized_PEC})} \\
 \hline \hline
Number of qubits & 2 & 21 & 92 & 20\\ 
\hline
Observables &$ZZ$, $ZI$ & $X^{\otimes n}$ & $Z_i, i \in \{0, \dots, 91\}$ & $Z_i, i \in \{0, \dots, 19\}$ \\
\hline
Number of twirls & 250 & 100 & 100 & {128 (learning) \\ 100.000 (PEC)} \\
\hline
Shots per twirl & 200 & 256 & 150 & {64 (learning) \\1 (PEC)}\\
\hline
{Even-depth \\learning layers for \\  symmetric model} & {$d$=\{2, 4, 6, \dots, 32\}} & $d = \{2, 4, 8 \}$ & $d = \{4, 12, 24 \}$ & $d = \{4, 10, 20, 40 \}$\\
\hline
{Model Pauli basis} & {restricted to Paulis \\ relevant for observable} &  {restricted to Paulis  \\ relevant for observable} & {all one- and \\ two-local Paulis} & {all one- and \\ two-local Paulis}\\
\hline
{Design matrix \\dimensions of self-\\ consistent model} & (9, 9) & (56, 46) & (9108, 2576) & (2349, 521) \\
\hline
{Model locality\\ assumption} & None & None & two-local & two-local \\
\hline
\end{tblr}
\caption{
Details for the  four experimental results, with increasing  complexity, presented in the same order as the main text. 
To transform the noise of both the learning circuits and the target circuits into Pauli noise, a number of logically equivalent circuits, referred to as ``twirls'' was implemented. For each twirl, the circuit was sampled with a given number of shots, where each shot lasted for approximately 1~millisecond. 
Depending on the target circuit, the number of even-layers used to infer the symmetric noise model ranged in depths from 8 to 40. 
Finally, the complexity of the noise model increased with the size of circuits, where for the two-qubit experiment the model Pauli basis was restricted to those that impacted the mitigated observables whereas the 20-qubit and 92-qubit experiment involved all one- and two-local Paulis. 
This also meant the design matrix for the self-consistent noise model grew not only in the number of rows (corresponding to the number of learned observables), but also the number of columns (corresponding to the noise parameters) which grow from 9 to 2576. 
We emphasize that the final two columns represent the \textit{scalable} approach, where the total number of learning circuits remains fixed no matter the size of the system; as discussed in the main text, this is because the noise is assumed to be two-local and thus long-range noise terms are not being considered. 
For examples of design matrices, see Fig.~\ref{sfig:design_2Q} for the 2Q column, and Table~\ref{tab:ghz} for the GHZ column. The larger design matrices are not shown, but can be reconstructed using code provided in~\cite{chen2026efficient}.
}
\label{tab:experiment_settings}
\end{center}
\end{table}

\subsection{Details on two-qubit experiments}
\label{sm:more_pair}
\begin{figure} 
	\centering
	\includegraphics[width=0.74\textwidth]{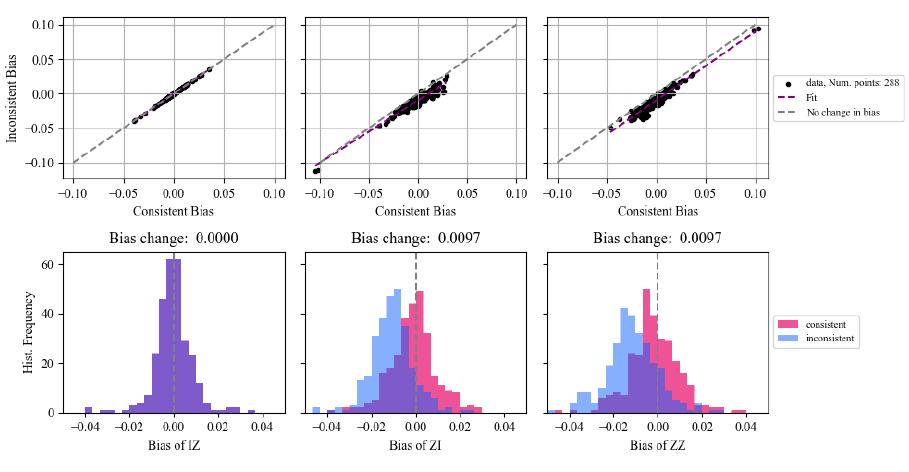} 
	\caption{\textbf{Two-qubit experiments across 27-qubit device, \textit{ibm\_auckland}.}
	(\textbf{top}) Comparison of bias in mitigated values of the consistent versus the inconsistent noise models for the $\langle IZ\rangle$,  $\langle ZI\rangle$, and $\langle ZZ\rangle$ observables.
	(\textbf{bottom}) Histogram distributions of the biases for the consistent (red) and the inconsistent (blue) noise models for the $\langle IZ\rangle$,  $\langle ZI\rangle$, and $\langle ZZ\rangle$ observables.
	}
	\label{sfig:two_qubit_histogram} 
\end{figure}

Whereas we discussed the details of the learning circuits earlier in Sec.~\ref{supp:constructing_design_matrix}, here we will focus on the two-qubit target circuit we examined. For this \textit{restricted} model experiment limited to $Z$-only observables, we prepared all learning circuits in the $|00\rangle$ state, but used that learned noise model to mitigate the outcomes of a circuit prepared in the $|11\rangle$ state. In this manner, by comparing the experimentally measured outcomes for the observables $\langle IZ \rangle$, $\langle ZI \rangle$, and $\langle ZZ \rangle$, against those predicted by the learned noise model, we were able to identify an improved bias, of up to $4\%$ for a single pair of qubits up to depth-32, and also an improved bias of $0.97\%$ across six qubits on \textit{ibm\_auckland}, a 27-qubit device (See~\ref{sfig:two_qubit_histogram}). This comprised of a total of 144 learning experiments, and 144 mitigation experiments taken over the course of a day.

\subsection{Details on GHZ-preparation experiments}
\label{sm:GHZexpt}
In Sec.~\ref{sec:GHZ} of the main text and above, we only focused our discussion on $n=5$ or $n=21$ GHZ states. However, the full data set involved all odd-sized GHZ states between $n=3$ and $n=21$, inclusive. We used 21 physical qubits on a line: $12, 17, 30, 31, 32, 36, 51, 50, 49, 55, 68, 69, 70, 74, 89, 88, 87, 93, 106, 105, 104$. We repeated the experiment a total of 7 times, interleaving the learning experiments and the target experiments over the course of a day (18 hours) immediately after the system \textit{ibm\_strasbourg} was calibrated. The standard deviations shown in Fig.~\ref{fig:6} were taken over the 7 experimental runs.

\subsection{Details on 92-qubit ring experiments}
\label{sm:ring_experiments}
\begin{figure} 
	\centering
	\includegraphics[width=0.74\textwidth]{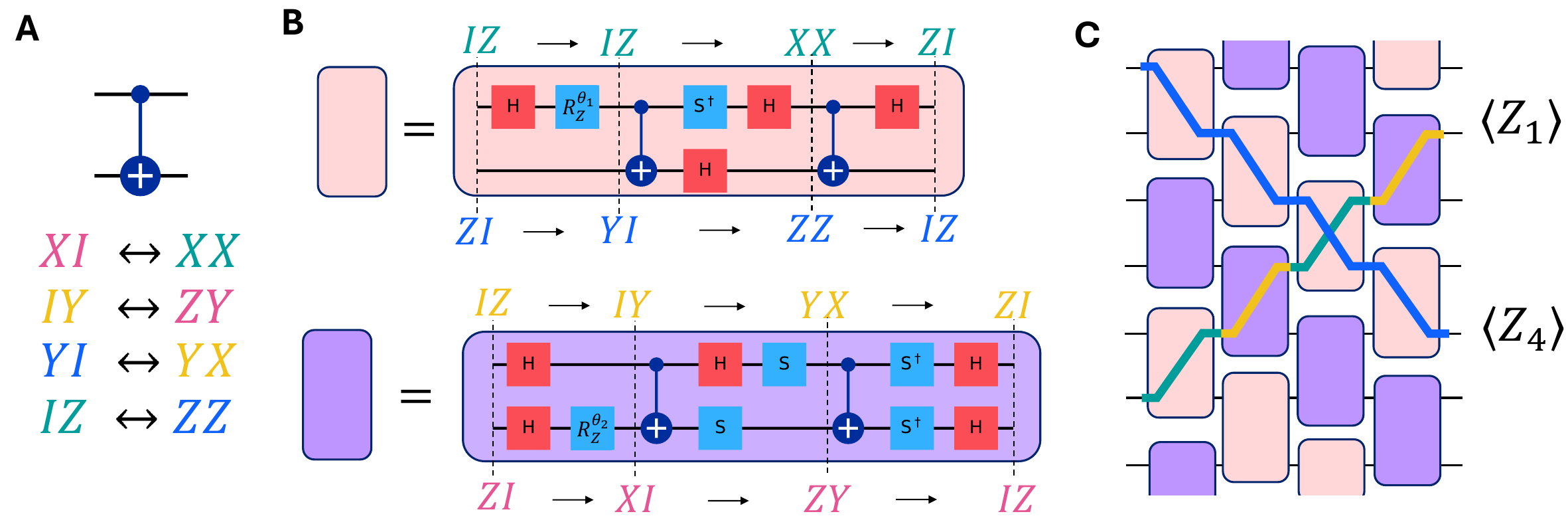} 
	\caption{\textbf{Circuit for ring experiments with weight-1 observables.}
	(\textbf{a}) A single CNOT gate has four conjugate Pauli fidelity pairs that change pattern under conjugation. 
	(\textbf{b}) Two qubit circuit blocks that transform $ZI \leftrightarrow IZ$, such that the noisy Pauli fidelities that affect each observable (before the unitary part of the gate) originate from different conjugate pairs, as indicated by color. 
    The indicated Pauli paths arise for the $R_z$ rotation angles $\theta_1 = \pi/2$ and $\theta_2 = -\pi/2$, which we use for the Clifford circuit experiments from Sec.~\ref{sec:ring}.
    (\textbf{c}) Arranging the two-qubit blocks from (b) in an alternating pattern, each single-qubit $Z$ observable propagates in a staircase shape, such that every degenerate fidelity from each CNOT gate shown in (a) is probed by one observable. 
	}
	\label{fig:methods_ring_circuit} 
\end{figure}

Here we detail the circuit and observables for the experiments presented in Sec.~\ref{sec:ring} of the main text.
This experiment was designed with the aim to probe all degenerate fidelity pairs of the CNOT gates of a one-dimensional, closed loop of qubits.
For a single CNOT gate, there are four degenerate cycles of conjugate Pauli pairs that change their pattern under conjugation with the CNOT gate. 
These are $IZ \leftrightarrow ZZ$, $XI \leftrightarrow XX$, $ZY \leftrightarrow IY$, and $YX \leftrightarrow YI$~\cite{chen2023learnability} (See Fig.~\ref{fig:methods_ring_circuit}a). 
We start by designing two different two-qubit blocks with two noisy CNOT gates each, such that all single-qubit $Z$ observables are sensitive to two Pauli fidelities that originate from different degenerate cycles. 
These two-qubit blocks are shown in Fig.~\ref{fig:methods_ring_circuit}a. 
Note that they shift the support of the $IZ$ observable to the other qubit.  
Hence, when arranging the blocks in the pattern shown in Fig~\ref{fig:methods_ring_circuit}C, each $Z$-observable propagates in a ``staircase''-like trajectory.
After four layers of the alternating pattern (eight total layers of CNOTs), every fidelity split of each participating CNOT connection is probed by one observable.
There are only two unique layers of CNOT gates, as the two-qubit blocks only differ in their single-qubit gate structure. 

With this construction, we ensure that the $Z$ observables in this experiment are maximally sensitive to asymmetries in the gate noise fidelities. 
Moreover, every observable is affected by the state preparation noise from one qubit and the measurement noise of a different qubit. 
However, traditional approaches of readout error mitigation (used for the ``symmetric model'' throughout this work) are sensitive to the state preparation error of the final, measured qubit~\cite{van2022model}.
Hence, these observables expose flaws in the symmetric model when state preparation errors are not uniform across the ring of qubits. 

The measured observables generally showed better agreement with the self-consistent noise model than with the symmetric noise model (See Fig.~\ref{fig:7}e).
However, for most observables, a residual bias remained also under the self-consistent noise model. 
This is an indication that there are error sources present in the device which even the consistent model does not accurately account for. 
Candidates for such errors could be leakage out of the qubit subspace, temporal drifts of the noise model between the learning circuits and the target circuits, remaining coherent errors, or non-nearest-neighbor correlated noise sources. 
Finally, few individual observables in Fig.~\ref{fig:7}E (e.g. qubit index 57) show a larger bias under the self-consistent model than the symmetric model. 
We note that this occurred predominantly when there were severe outliers in the individual Pauli or SPAM fidelities affecting the respective observables. This could be caused, e.g., by the presence of two-level systems (TLS). 
These lead to strong fluctuations in the noise parameters on short time scales, to which the self-consistent learning protocol is particularly vulnerable due its dependence on depth-one circuits. 
We thus expect that our learning protocol will further benefit from recent techniques to stabilize the noise~\cite{kim2024error}.

\subsection{Details on PEC experiment}
\label{sm:pec_circuits}

We built on the benchmark circuits introduced in Sec.~\ref{sm:ring_experiments} for our proof-of-principle demonstration of gauge-optimized PEC presented in Sec.~\ref{sec:gauge-optimized_PEC}. 
Specifically, we used versions of the circuits presented in Fig.~\ref{fig:methods_ring_circuit} obtained by choosing $\theta_1=-\pi/3$ and $\theta_2 = 2\pi/3$.
In this setting, the Pauli paths of the measured single-qubit $Z$ observables become more complicated than in the Clifford case; they generally split into two branches with every two-qubit block, but still probe the degenerate fidelity pairs of each CNOT and are sensitive to differences in SPAM errors. 
Instead of running the circuits on a closed loop of qubits with periodic boundary conditions as in Sec.~\ref{sec:ring}, for this experiment we chose to run the experiments on an open-ended 1d string of qubits to further optimize the error rates of the chosen qubit layout. 
In this setting, the two-qubit blocks that are only shown halfway in Fig.~\ref{fig:methods_ring_circuit}(c) were removed from the circuit. 
Hence, at a circuit size of 20 qubits, the two different layers of parallel CNOT gates have 10 and 9 gates, respectively. 
At a depth of 10 layers of alternating blocks, this led to 190 CNOT gates in total as shown in Fig.~\ref{fig:PEC_results}(b).

\section{Gauge optimization}
\subsection{Difference in $\gamma$ between two strategies}
\label{sm:gauge_opt}
Our initial na\"{i}ve attempt at optimizing the gauge involved using the Moore-Penrose pseudo-inverse to recover $\bm{\hat{r}}_0=\bm{F'}^+ \bm{b}$, which is effectively a least-squares minimization problem of the form $\epsilon = \left\lVert \bm{F'} \bm{r}_0 - \bm{b}\right\rVert_2$~\cite{2020SciPy-NMeth}. This is followed by a second optimization step:
\begin{equation}
	\min_{\bm{\eta}\in\mathbb{R}^n} 
	\left\{
		\sum_{i=1}^{24n} 
		\text{max} 
			\left( 
				\left[\bm{A}_{r}^{\tau}(\bm{r}_0 + \bm{S}^{\dagger} \bm{\eta})\right]^{(i)}, 
				0
			\right) 
	\right\}
	\label{eq:gauge_opt_2step} %
\end{equation}
where the $k$th column of the matrix $\bm{S}^{\dagger}$ is one of $n$ vectors $\bm{y}_k$ in the nullspace of the design matrix. In other words, $\bm{S}^{\dagger}$ converts each gauge parameter $\bm{\eta}_k$ into a vector in the noise parameter space $\mathcal{X}$ such that the residual errors $\epsilon$ between the model and measured outcomes remains unchanged: $\bm{F'} \bm{\hat{r}}_0 - \bm{b} = \bm{F'} (\bm{\hat{r}}_0 + \bm{S}^{\dagger}\bm{\eta}) - \bm{b}$. The summand in the optimization problem limits all $24n$ elements of the $\bm{\tau}$ vector to be positive definite before being summed, element-wise, together. The optimized $\bm{\hat{r}}_{*}^{\text{two-step}}$ is then an offset of $\bm{\hat{r}}_0$: $\bm{\hat{r}}_0 + \bm{S}^{\dagger} \bm{\eta}^{*}$, where finding $\bm{\eta}^{*}$ involves solving the convex optimization problem in Eq.~\eqref{eq:gauge_opt_2step}. 

\begin{figure} 
	\centering
	\includegraphics[width=0.5\textwidth]{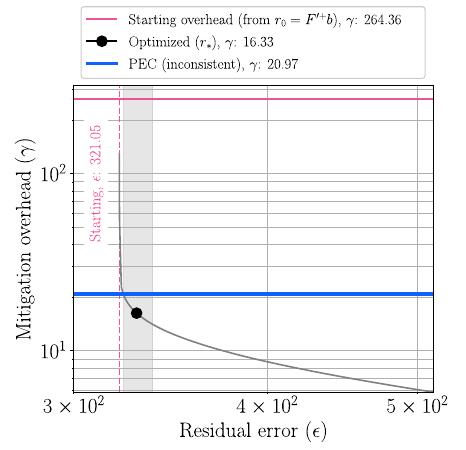} 
	\caption{Gauge optimization for the noise model of the 92-qubit experiments (Sec.~\ref{sec:ring}), where the target residual error $\epsilon = | F'\bm r – \bm b|$ can be varied. The black point was chosen as the optimal trade-off between residual error and overhead minimization, with the shaded gray region indicating the range of $\bm r$ values we chose from.}
	\label{sfig:gauge_opt} 
\end{figure}

However, benchmarking on the noise model data from the 92-qubit experiment from Sec.~\ref{sec:ring}, we found that this two-step approach, which started with a $\gamma\approx264$ without any optimization step (no steps beyond calculating $\bm{\hat{r}}_0=\bm{F'}^+ \bm{b}$), yielded much higher overheads than even the $\gamma$ inferred from a non-negative least-squares fit based on previous approaches, where $\gamma \approx 20.97$~\cite{van2023probabilistic}. Thus, the optimization procedure we outlined in Sec.~\ref{sec:gauge_opt} yielded a slightly higher residual error ($\epsilon=329.39$) but at a significantly lower $\gamma=16.33$ (See Fig.~\ref{sfig:gauge_opt}).

\subsection{Gauge-optimized model for PEC}
\label{sm:PEC_shot_noise}
\begin{figure} 
	\centering
	\includegraphics[width=0.5\textwidth]{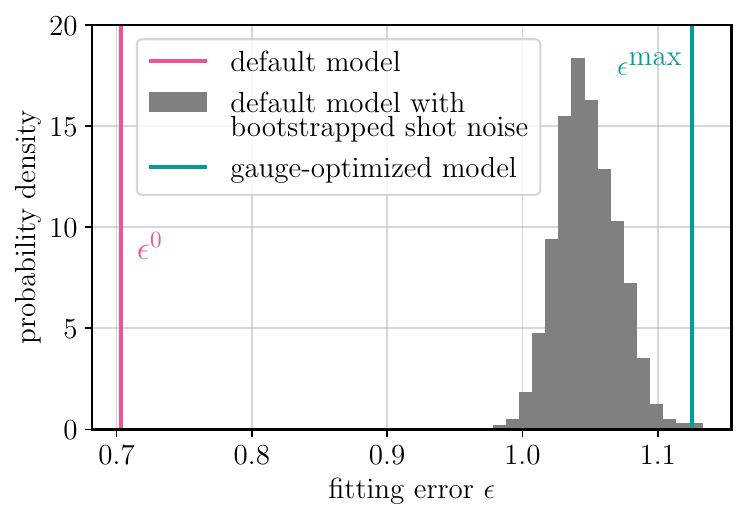} 
	\caption{Illustration of the choice of the residual error $\epsilon_\text{max}$ for the gauge-optimized PEC experiment presented in Sec.~\ref{sec:gauge-optimized_PEC}, where $\epsilon_\text{max}$ was chosen such that it lies just within the observed fitting errors of the original model when accounting for shot noise in the learning data through bootstrapping.}
	\label{sfig:model_error_bootstraps} 
\end{figure}

The above analysis illustrates that there is a bias-variance tradeoff when choosing the gauge such that the PEC overhead $\gamma$ is minimized. 
As shown in Fig.~\ref{sfig:gauge_opt}, a small concession in the fitting error $\epsilon$ (which may introduce a bias due to a worse fit of the noise model to the learning data) can significantly reduce the variance of a PEC estimator. 
For our demonstration of gauge-optimized PEC in Sec.~\ref{sec:gauge-optimized_PEC}, we navigated this tradeoff by exploring a region of possible $\epsilon$ that are within the shot noise fluctuations of the learning data, as detailed in the following.

First, we note that the vector $\hat{\bm b}$ of measured log-fidelities in Eq.~\eqref{eq:design_matrix} is subject to shot noise on the individual expectation values $\tilde o_j$. 
Let us denote the statistical errors on these expectation values as $\delta \tilde o_j$. 
We obtained the ``default'' gauge $\hat{\bm{r}}^\text{0}$ of the learned noise model by solving Eq.~\eqref{eq:gauge_opt_1step} directly, where the sum over layers was taken such that each mitigated layer enters the expression as many times as it occurs in the circuit, i.e., once for the state preparation layer and ten times for each of the two CNOT layers. This way, the cost function directly represents the $\gamma$-value of the full circuit, for which we obtain $\gamma^{\text{0}} = 48.7$. 
Note that we mitigate readout errors in post-processing by dividing the PEC estimators through the readout fidelities corresponding to the measured observables, in the spirit of the TREX method.

The resulting fitting error $\epsilon^\text{0}$ establishes a baseline shown as the leftmost vertical line in Fig.~\ref{sfig:model_error_bootstraps}.
Next, we computed bootstrapped vectors $\hat{\bm b}_{\text{bs}}$ by sampling new sets of values for $\tilde o_j$ drawing from appropriate Bernoulli statistics (since all observables are Paulis with possible measurement outcomes of $\pm1$) with mean $\tilde o_j$ and standard deviation $\delta \tilde o_j$. 
We repeated this process 1000 times to obtain fittings errors $|| {F'} \hat{\bm{r}}^\text{0} - \hat{\bm{b}}_\text{bs} ||$. The distribution of these errors is shown as a histogram in Fig.~\ref{sfig:model_error_bootstraps}. 
For the gauge-optimized model, we chose the maximum allowed fitting error as $\epsilon^\text{max} = 1.6\epsilon^0$, which lies within the tail of the distribution of the bootstrapped values, see Fig.~\ref{sfig:model_error_bootstraps}. 
This reduced the $\gamma$ value to $3.7$, which corresponds to an improvement in sampling overhead from $2372$ for the default gauge to only 13.7, an improvement by a factor of 173.

\subsection{Computational efficiency of gauge optimization}\label{sm:convex}

We now discuss the computational efficiency of the gauge optimization problem.
For convenience, we restate the gauge optimization problem:
\begin{equation}
	\begin{aligned}
	&\min_{\hat{\bm{r}}}  
	\left\{
		\sum_{a\in\mc K,~\text{layer}}
		\text{max} 
		\left(\hat\tau_a^{\text{layer}}, 
			0
		\right) 
	\right\},
	\\
	&~~\text{s.t.}~  \left\lVert {F'} \hat{\bm{r}} - \hat{\bm{b}} \right\rVert_2 \le \epsilon.
	\end{aligned}
\end{equation}
Here, the summation is over all layers (including both SPAM and gates) of noise and all Pauli operators consistent with our quasi-local noise ansatz (described by $\mc K$). For each layer, $\hat{\bm\tau}$ can be obtained from $\hat{\bm r}$ via the linear transformation defined in Eq.~\eqref{eq:r2tau_theory}. We will assume there are polynomially many layers and $\mc K$ contains polynomially many Paulis (which is a reasonable requirement for a parameter-efficient noise ansatz, say nearest-neighbor two-local).

\medskip

The above gauge optimization is a convex optimization problem. More specifically, it is a second-order cone programming (SOCP) problem. Indeed, it can be rewritten in the following equivalent form:
\begin{equation}
    \begin{aligned}
        \min_{\hat{\bm r},\bm s} &\sum_{i=1}^{|\bm\tau|}s_i,
        \\\mr{s.t.}&~ s_i\ge (L\hat{\bm r})_i,
        \\&~s_i\ge0,
        \\&~\|F'\hat{\bm r} - \hat{\bm b}\|_2\le\varepsilon.
    \end{aligned}
\end{equation}

Here, we collect all $\tau_a^{\mr{layer}}$ into one column vector $\bm\tau$ of length $|\bm\tau|$, which is completely determined by $\bm{\hat r}$ via a linear transformation we denote by $\bm\tau =L{\hat{\bm r}}$. The equivalence between the above two optimization problems is obvious, and the latter is in a standard SOCP form (which consists of linear target functions and affine or 2-norm restrictions). Therefore, this problem is provably solvable to any given precision in polynomial time using the interior-point method~\cite{lobo1998applications}.

In our experiments, we use CVXPY~\cite{cvxpy}, a standard python-based convex optimization toolbox, to solve the gauge optimization problem. For our $92$-qubit nearest-neighbor noise model experiments, the gauge optimization is solved within a few seconds on laptop, justifying its computational efficiency in practice.

\end{document}